\documentclass[aps,prd,twocolumn,superscriptaddress,nofootinbib,amsmath, amssymb]{revtex4-2}
\renewcommand{\vec}[1]{\mathbf{#1}}

\usepackage[dvipsnames]{xcolor}
\usepackage[normalem]{ulem}
\usepackage{graphicx}
\usepackage{ifthen}
\usepackage{color}  

\definecolor{americanrose}{rgb}{1.0, 0.01, 0.24}
\definecolor{bananayellow}{rgb}{1.0, 0.88, 0.21}
\definecolor{britishracinggreen}{rgb}{0.0, 0.26, 0.15}
\definecolor{ao(english)}{rgb}{0.0, 0.5, 0.0}

\usepackage{hyperref}
\RequirePackage{hyperref}

\newcommand{\myparallel}{{\mkern3mu\vphantom{\perp}\vrule 
depth 0pt\mkern2mu\vrule depth 0pt\mkern3mu}}

\usepackage[mathscr]{euscript}

\DeclareSymbolFont{rsfs}{U}{rsfs}{m}{n}
\DeclareSymbolFontAlphabet{\mathscrsfs}{rsfs}
\newcommand{\myd}{\mathscrsfs{D}}
  
\newboolean{includeFigures}
\setboolean{includeFigures}{false} % Change this to false to exclude figures
\newcommand{\conditionalFigure}[1]{
    \ifthenelse{\boolean{includeFigures}}{#1}{}
}

\begin{document}
\title{A local diagnostic program for unitary evolution in general space-times}

\author{Ka Hei Choi} \email{K.Choi@physik.uni-muenchen.de}%Lines break automatically or can be forced with \\
\affiliation{Arnold Sommerfeld Center for Theoretical Physics, Theresienstra{\ss}e 37, 80333 M\"unchen, Germany\\}

\author{Stefan Hofmann}
\email{Stefan.Hofmann@physik.uni-muenchen.de}%Lines break automatically or can be forced with \\
\affiliation{Arnold Sommerfeld Center for Theoretical Physics, Theresienstra{\ss}e 37, 80333 M\"unchen, Germany\\}

\author{Marc Schneider}
\email{mschneid@sissa.it }%Lines break automatically or can be forced with \\
\affiliation{
SISSA, Via Bonomea 265, 34136 Trieste, Italy}
\affiliation{INFN Sezione di Trieste, Via Valerio 2, 34127 Trieste, Italy}
\affiliation{IFPU - Institute for Fundamental Physics of the Universe, Via Beirut 2, 34014 Trieste, Italy}

\date{\today}

\begin{abstract}
We present a local framework for investigating non-unitary evolution groups pertinent to effective field theories in general semi-classical spacetimes. Our approach is based on a rigorous local stability analysis of the algebra of observables and solely employs geometric concepts in the functional representation of quantum field theory. In this representation, it is possible to construct infinitely many self-adjoint extensions of the canonical momentum field at the kinematic level, and by the usual functional calculus arguments this holds for the Hamiltonian, as well. However, these self-adjoint domains have only the trivial wave functional in common with the solution space of the functional Schrödinger equation. This is related to the existence of boundaries in configuration field space which can be penetrated by the probability flux, causing probability to leak into regions in configuration field space that require a more fundamental description.  As a consequence the evolution admits no unitary representation. Instead, in the absence of ghosts, the evolution is represented by contractive semi-groups in the semiclassical approximation. This allows to quantify the unitarity loss and, in turn, to assess the quality of the semi-classical approximation. We perform numerical experiments based on our formal investigations to determine regions in cosmological spacetimes where the semiclassical approximation breaks down for free quantum fields. 
%\vspace{4em}
\end{abstract}

\maketitle

\section{Introduction}

Once quantum fluctuations of spacetime are excited, the semiclassical approximation for quantum fields in curved spacetimes implies a non-unitary evolution, because the approximation assumes an inert classical motion for the background geometry \cite{ Vilenkin:1988yd, Kuo:1993if, matsui2020quantum, massar1999unitary, anderson2003linear}.
Unitarity violations are concomitant with sources or sinks for probabilities. Assuming the quantum theory is consistent (in Minkowski spacetime), sources violating probability conservation cannot exist while probability sinks are allowed, implying that the spectrum of the Hamiltonian contains eigenvalues with a negative imaginary part.
This can be interpreted as a measure of the extent to which the semiclassical approximation is invalidated.
Of course, probability sinks have only operational consequences, provided that they involve detectable unitarity violations,
which, in turn, translate into the breakdown of the semiclassical description.

The corresponding evolution operators are a contractive representation of the time translation group. In the spirit of Stone’s theorem, which establishes a one-to-one correspondence between self-adjoint operators and certain one-parameter unitary groups, contractive representations enjoy accretive generators \cite{ree75}. While this grants the semiclassical approximation predictive power and is a nice feature at the technical level, it also qualifies the underlying Hamiltonian as non-observable if regions of spacetime are considered that contain probability sinks.

Any description of a system in curved spacetimes within the semiclassical framework can, therefore, at best be an effective field theory \cite{Burgess:2009ea}. This holds, in particular, for free quantum fields coupled to dynamical background geometries that are assumed to be inert against quantum fluctuations. This assumption might be violated in spacetime regions which support fluctuations of observables that violate the corresponding mean field. At the operational level, any observer must single out a set (algebra) of observables which they intend to measure in certain regions of spacetime. Statements concerning stability can only be made relative to this set. The validity of the semiclassical approximation depends on the probability to excite quanta beyond a boundary in field configuration space. The boundary field configuration relates to a geometric argument based on semi-norms concerning the stability of background observables.

\begin{figure}[b!]
    \centering
    \includegraphics[width=0.75\linewidth]{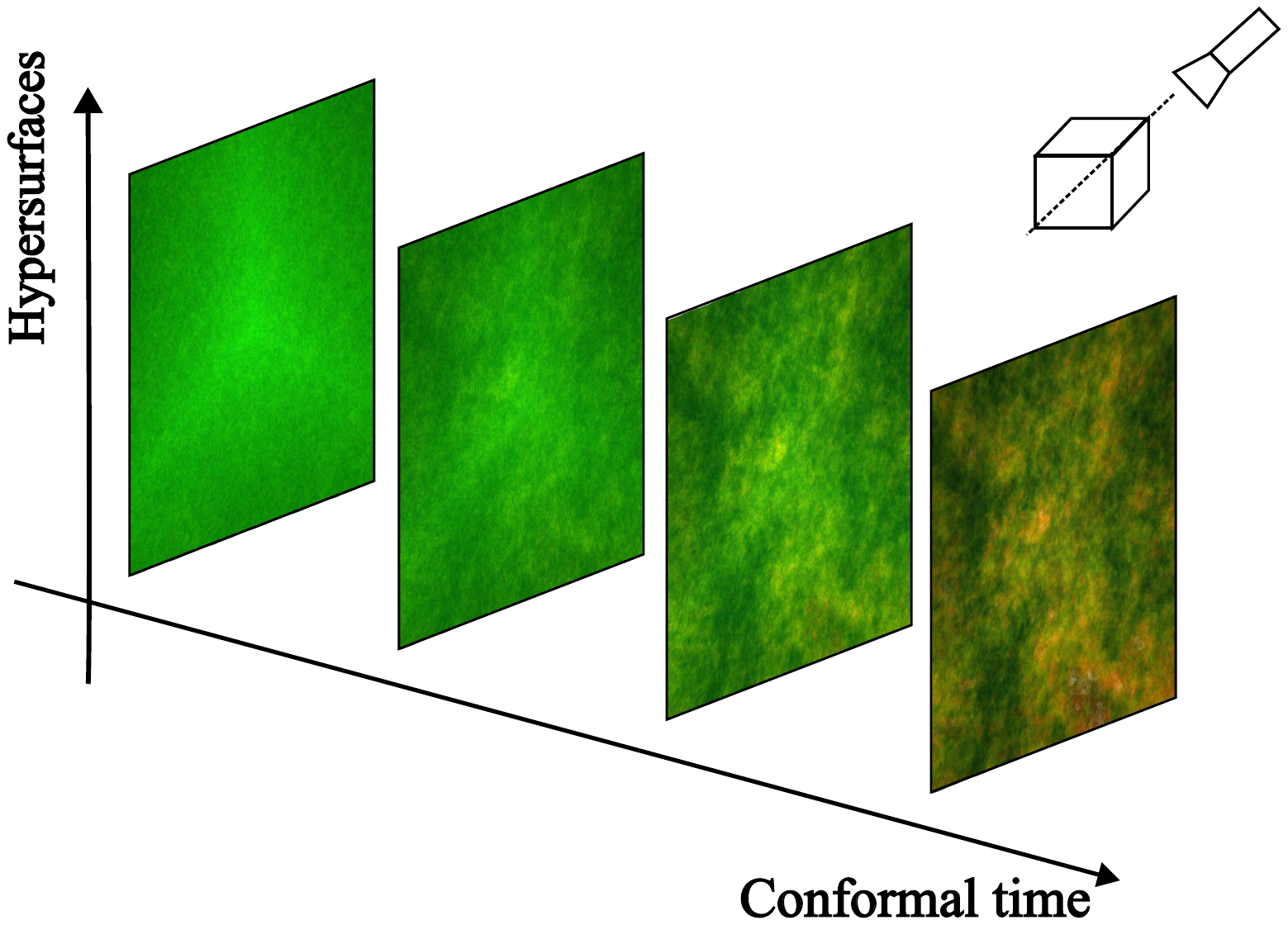}
    \caption{Simulation of free quantum fluctuations in de Sitter cosmology at different times. Shown are two-dimensional spatial slices corresponding to the perspective indicated by the camera depicted in the right top corner at different times. The slice through the initial hypersurface is to the far left. The colouring represents the distribution of fluctuation amplitudes. From left to right pattern formation is apparent. For a quantitative analysis of these simulations we refer the reader to Section \ref{sec:cst}.}
    \label{fig:intro_fig}
\end{figure}

The existence of such a boundary is crucial as there are no asymptotic conditions imposed on the domains of the functional integrals and, instead, observables have to be evaluated on finite boundaries. The boundary field configurations also impact the spectral analysis of field operators. As we show in the main body of the text, the canonical pair consisting of a real scalar field and its conjugated momentum field enjoy self-adjoint extensions on certain Hilbert spaces, hence they constitute the basic observables. However, the notion of basic observables is only meaningful at the kinematical level: the solution space of Schr\"odinger wave functionals might not intersect the Hilbert space on which the canonical pair admits a self-adjoint realization. From this point of view, quantisation (of polynomial observables) clashes with the dynamical content of the theory.

The functional Schr\"odinger picture of quantum field theory (cf. \cite{Eboli:1988qi, hatfield2018quantum, Halliwell:1991ef, Long:1996wf, Traschen:1990sw,  
guth1985quantum, PhysRevD.49.2769, Floreanini:1986tq, Guven:1987bx, 
Hofmann:2015xga, Hofmann:2016vix, Eglseer:2017kcs, Hofmann:2019dqu,  
Callan:1994py, Berges:2017hne, 
Corichi:2002ir, torre1998quantum, 
Alonso:2023ien,Alonso:2023kvy, Alonso:2024ajv} for details) allows to address the questions raised above, in particular, to determine the domain of validity associated with the semiclassical approximation and its concomitant unitarity violation. In the absence of ghosts, unitarity violation is tied to a significant probability flux across the boundary in field configuration space. Whether the violation of probability conservation is acceptable depends on the resolution with which observables are measured. If unitarity violations are acceptable within the spacetime region associated with the measurement process, then contractive representations generalize unitary evolution groups and grant predictive power even when the probabilistic framework underlying effective quantum theories becomes challenged.

The formal arguments presented in the main body of the text are applied to simulations employing Gaussian random fields. Based on these numerical experiments, we demonstrate the potential of the formalism by studying free fields in cosmological spacetimes, such as contracting radiation dominated universes evolving towards a future singularity and the asymptotic past of de Sitter universes. The simulations reveal regions in these spacetimes, where the semiclassical approximation fails relative to a geometric stability criterion on the algebra of observables.

The article is organized as follows: In Sec. \ref{sec:ecs} the kinematic aspects of effective field theories in curved spacetime are discussed in the Schrödinger representation, including preliminaries and new topics such as effective configuration field spaces and stability analysis based on geometric and functional methods, spectral properties of functional operators in the presence of configuration field boundaries, the impact of these boundaries on self-adjoint extensions of the functional momentum operator and the functional Hamilton operator at the kinematic level. Sec. \ref{sec:da} confronts the results obtained in the previous section with dynamic aspects. The following Sec. deepens the preceding discussion and takes a fresh look at contractive evolution semi-groups. This concludes the formal considerations of the article. In Sec. \ref{sec:cst} we apply our formal results to study effective field theories in cosmological spacetimes and domains of validity of the semiclassical approximation by employing numerical experiments. The conclusion of the present work is presented in Sec. \ref{sec:conclude} and followed by two appendices, App. \ref{app:notation} and \ref{app:nt}, concerning the notation employed in this article and some technical aspects of the numerical experiments.

\section{Kinematic Aspects}
\label{sec:ka}
In this section, we provide a brief overview on the kinematic aspects of the functional Schr\"odinger representation in quantum field theory. For simplicity, we consider only globally hyperbolic spacetimes
$(\mathcal{M},g)$, where $\mathcal{M}$ is a connected four-dimensional, 
smooth, Hausdorff manifold, and $g$ is a Lorentzian metric. 
The manifold $\mathcal{M}$ is diffeomorphic to $I\times\Sigma$
with $I\subseteq\mathbb{R}$ \cite{ger70}, and it foliates into 
hypersurfaces $(\Sigma_t)_{t\in\mathbb{R}}$.
Relative to this foliation, the metric is given by 
\begin{eqnarray}\label{admmetric}
    g
    &=&
    \left(
        -N_\perp^{\;\; 2}+N_\myparallel (N^\myparallel)
    \right) \mathrm{d}t\otimes\mathrm{d}t 
    +
    2 N_\myparallel \otimes \mathrm{d}t
    +
    q
\; ,
\end{eqnarray}
where $N_\perp$ denotes the lapse function, $N_\myparallel$ 
the dual shift vector, and $q$ is a Riemannian metric on 
$\Sigma_t$.

For $t\in I$, let $\mathcal{C}_t$ be a real vector bundle 
over the hypersurface $\Sigma_t$, given by the quadruple
$(C_t,\pi_t,\Sigma_t,(V_{p})_{p\in\Sigma_t})$, 
where $C_t$ denotes the total space, $\pi_t$ is the 
bundle projection of $C_t$ onto the base $\Sigma_t$, 
and $V_p$ is a real vector space homeomorphic to the 
fibre $\pi^{-1}(p)$. In the following, we consider only 
trivial bundles of rank one over $\Sigma_t$ and refer 
to $\mathcal{C}_t$ as the configuration bundle of 
instantaneous field configurations. 
Configuration fields $\phi_t$ are smooth sections of $\mathcal{C}_t$, 
collected in the vector space $\Gamma(\mathcal{C}_t)$. 
Of particular interest is the space 
$\Gamma_\mathrm{c}(\mathcal{C}_t)\subset\Gamma(\mathcal{C}_t)$
of compactly supported smooth sections of the configuration bundle
$\mathcal{C}_t$. 

Let $(\Gamma(\mathcal{C}_t),\myd\phi)$ denote a formal measure space.
Consider the complex vector space 
$\mathcal{L}^2(\Gamma(\mathcal{C}_t),\myd\phi)$
of measurable complex-valued 
wave functionals $\Psi_t$ in $\Gamma^*(\mathcal{C}_t)$, 
whose modulus is square integrable with respect to the formal Lebesgue
measure $\myd\phi$.
Let 
$\mathfrak{N}(\Gamma(\mathcal{C}_t),\myd\phi)\subset 
\mathcal{L}^2(\Gamma(\mathcal{C}_t),\myd\phi)$
be the subspace consisting of wave functionals that vanish 
$\myd\phi$-almost everywhere. The quotient space 
$L^2(\Gamma(\mathcal{C}_t),\mathcal{D}\phi)$ $:=$
$\mathcal{L}^2(\Gamma(\mathcal{C}_t),\myd\phi)
\setminus\mathfrak{N}(\Gamma(\mathcal{C}_t),\myd\phi)$
can be equipped with a norm 
$\|\Psi_t\|=\mathbb{E}^{1/2}(\mathrm{id}_{\Gamma^*(\mathcal{C}_t)} \, ; \, \Psi_t)$, where 
\begin{eqnarray}
    \mathbb{E} \left(\mathcal{O}\, ; \, \Psi_t \right)
    :=
        \int_{\Gamma(\mathcal{C}_t)}
        \myd\phi \; 
        \overline{\Psi_t}(\phi) \, \mathcal{O} \, \Psi_t(\phi):=\langle\mathcal{O}\rangle_{\Psi_t}
\end{eqnarray}
for any bounded operator $\mathcal{O}$ on 
$L^2(\Gamma(\mathcal{C}_t),\myd\phi)$. 
Note that while $\|\cdot\|$ is a formal norm on 
$L^2(\Gamma(\mathcal{C}_t),\myd\phi)$, 
it is merely a semi-norm on 
$\mathcal{L}^2(\Gamma(\mathcal{C}_t),\myd\phi)$.

Concerning the underlying probabilistic framework (refraining from 
topological issues): if $\mathcal{U}$
is a measurable subset of $\Gamma(\mathcal{C}_t)$, let
$\mathcal{X}_\mathcal{U}:\Gamma(\mathcal{C}_t)\rightarrow \{0,1\}$
denote the functional 
$\phi\mapsto (\phi, \mathrm{id}_{\Gamma(\mathcal{C}_t)}|_\mathcal{U} \cdot \phi)
/(\phi, \mathrm{id}_{\Gamma(\mathcal{C}_t)} \cdot \phi)$, then 
$\|\mathcal{X}_\mathcal{U} \Psi_t\|^2$ is the probability for
the instantaneous configuration fields in $\mathcal{U}$ to populate 
the hypersurface $\Sigma_t$. The brackets $(\cdot,\cdot)$ denote the inner product associated with $\Gamma(\mathcal{C}_t)$. This interpretation assumes 
that $\Psi_t$ has been normalized to unity on the initial hypersurface. 

The first elementary observable we consider is the smeared 
configuration field operator $\Phi(f)$, where $f$ 
is an instantaneous field configuration of compact support, 
$f\in \Gamma_\mathrm{c}(\mathcal{C}_t)$. The domain of $\Phi(f)$
is given by all wave functionals in $L^2(\Gamma(\mathcal{C}_t),\myd\phi)$ 
such that $\Phi(f) \Psi_t$ is in $L^2(\Gamma(\mathcal{C}_t),\myd\phi)$
for all admissible smearing functions.
For wave functionals in this domain, 
$\langle\Phi(f)\rangle_{\Psi_t}:=\langle\phi(f)\rangle_{\Psi_t}$. In other words, 
the smeared configuration field operator acts as a multiplication 
operator on its domain in $\Gamma^*(\mathcal{C}_t)$.
A measure for the probabilistic scatter of instantaneous 
configuration fields around the expectation value 
with respect to the wave functional $\Psi_t$ is 
$
    \mathbb{E}(\mathrm{id}_{\Gamma(\mathcal{C}_t)}
    \, ; \,
    (\Phi(f) - \phi(f) \mathrm{id}_{\Gamma^*(\mathcal{C}_t)}) \Psi_t)
$.

The second elementary observable, the smeared momentum field operator $\Pi(f)$ conjugated to $\Phi(f)$
is defined as the functional generalization 
of a directional derivative\footnote{Let $f\in\mathcal{C}_c^\infty(\mathbb{R},\mathcal{M})$ be a test function of compact support, the functional derivative can be thought of as a generalized directional derivative $\delta_\phi F(\phi):=\lim_{\lambda\to0}\lambda^{-1}F(\phi+\lambda f)$ where the generalization to multiple applications of $\delta_\phi$ follows immediately \cite{fredenhagen2015perturbative}.}
such that the canonical commutation relation
$[\Phi(f_1),\Pi(f_2)]=\mathrm{i}(f_1,f_2)$ holds.
We refrain from giving a precise characterization of the 
domain of $\Pi(f)$ for now, but it certainly encompasses those 
wave functionals $\Psi_t$ in $L^2(\Gamma(\mathcal{C}_t),\myd\phi)$ with
$\Pi(f)\Psi_t\in L^2(\Gamma(\mathcal{C}_t),\myd\phi)$ for all 
smooth smearing functions $f$ of compact support on $\Sigma_t$. The matter of the domain will be the subject of a later discussion in Sec. \ref{sec:fmo}. 
The canonical commutation relations imply Heisenberg's uncertainty 
relation.
From these elementary observables others can be formed 
using a natural generalization of functional calculus 
as characterized in the spectral theorems of unbounded operators. 

This concludes our brief review of the kinematic aspects 
underlying the Schr\"odinger picture of quantum field theory 
in Lorentzian spacetimes.

\subsection{Effective Configuration Spaces}\label{sec:ecs}
So far we considered formal measure spaces $(\Gamma(\mathcal{C}_t), \myd \phi)$
with $t$ in some interval $I\subset\mathbb{R}$ as configuration spaces 
and wave functionals $\Psi_t$ in the dual vector bundle $\Gamma^*(\mathcal{C}_t)$
satisfying $\mathbb{E}(\mathrm{id}_{\Gamma^*(\mathcal{C}_t)} ; \Psi_t)<\infty$.
Clearly, $(\Gamma(\mathcal{C}_t), \myd \phi)$ contains some field configurations 
that are inadmissible for a semi-classical treatment. 
Therefore, in this section, we construct an admissible configuration space. 

In classical field theory, 
a local observable $\mathcal{O}$ is a smooth section in 
$\Gamma^*(\mathcal{C}_t)$ such that for each background 
configuration $\phi_0\in \Gamma(\mathcal{C}_t)$
there exists a neighborhood $U_0$ in $\Gamma(\mathcal{C}_t)$
in which for all fluctuations $\phi$ relative to $\phi_0$ in $U_0$
we have $\mathcal{O} = \omega \circ j^k$, 
where $j^k_{(t,p)}(\phi)$ denotes the $k^{\rm th}$ jet prolongation 
of $\phi$ at $(t,p)\in I\times \Sigma_t$, and $\omega$
is a functional on the jet bundle $J^k\Gamma(\mathcal{C}_t)$, 
where $k\in\mathbb{N}$. 
Note that we hide any dependence on the background geometry
for the ease of notation. 
The canonical quantization prescription
allows to map any local observable to a (bounded) operator 
on $L^2(\Gamma(\mathcal{C}_t),\myd \phi)$, denoted 
by the same symbol for simplicity. 
Note that the definition of local observables in classical field 
theory is adapted to a covariant framework, while the quantization prescription 
in the functional Schr\"odinger picture
requires a foliation of spacetime into hypersurfaces\footnote{
From this perspective it might have been more appealing to 
consider local functionals on the momentum phase space 
of a given classical field theory. We refrain to follow this logic in order
to keep the presentation concise.
}.

In the following, we assume that the
background configuration $\phi_0$ is in 
$\mathrm{ker}(P)$, where $P$
is the hyperbolic wave operator associated with the free theory. 

We choose a set $\mathcal{A}_0$ 
of finitely many local quantum observables  
with support in $U_0\subset \Gamma(\mathcal{C}_t)$.
Our first objective is to characterize $U_0$ further.
Any quantum observable in $\mathcal{A}_0$
can be represented as
$\mathcal{O}_0 \, {\mathrm{id}}_{U_0}$
$+$ $\mathcal{O}(\phi)$ in $U_0$, where 
$\mathcal{O}_0$ denotes the corresponding classical observable 
evaluated at the background configuration. 
Let
$d(\mathcal{O}_0)$
denote the length dimension of $\mathcal{O}_0$, 
and let $\ell$ be a short-distance cut-off
below which a fundamental description is eventually required\footnote{
In fact, there might be an ordered set $\{\ell_a\}_{a\in(n)}$
of cut-offs such that the description of a finite number 
of observables in the effective theory with cut-off $\ell_a$
is contained in the description labeled with 
cut-off $\ell_{a+1} \leq \ell_a$.}. 
We consider
$\mathcal{O}_\ell := \ell^{d(\mathcal{O}_0)}$ 
as the short-distance limit of the quantity corresponding to $\mathcal{O}$, 
which allows to introduce the dimensionless ratio
\begin{eqnarray}
    \mathcal{R}(\mathcal{O})
    :=
    \mathcal{O/\mathcal{O}}_\ell \; .
\end{eqnarray}
Given a wave functional $\Psi_t\in L^2(\Gamma(\mathcal{C}_t),\myd\phi)$, 
we introduce the filtering semi-norm
\begin{eqnarray}
\label{sns}
    N_{\phi_0,\Psi_t}\left(\mathcal{R}(\mathcal{O})\right)
    :=
    \sup\limits_{U_0\subset \Gamma(\mathcal{C}_t)}
    \left|
        \mathbb{E}^{-s}\left(\mathcal{X}_{U_0}\cdot 
        \mathcal{R}(\mathcal{O});\Psi_t\right)
    \right|
\end{eqnarray}
where $s:= \mathrm{sgn}(d(\mathcal{O}_0))$; for any background configuration $\phi_0\in\mathrm{ker}(P)$, $t$ in the interval $I$, and require 
\begin{eqnarray}
\label{cond_sd}
    \max_{\mathcal{O}\in\mathcal{A}_0}
    \Big(N_{\phi_0,\Psi_t}
    \big(\mathcal{R}(\mathcal{O})\big)\Big)
    < 1 \;,
\end{eqnarray}
for any sensible fluctuation. If $L_\mathcal{O}^{d(\mathcal O_0)}$ denotes the length scale characterizing 
$\mathcal{O}$, then $N_{\phi_0,\Psi_t}\left(\mathcal{R}(\mathcal{O})\right)$ 
scales as $(\ell/L_\mathcal{O})^{|d(\mathcal{O}_0)|}$
for either sign of $d(\mathcal{O}_0)$.
Any observable in $\mathcal{A}_0$ violating this bound requires
a description characterized by a short-distance 
cut-off $\ell_1 < \ell$, including $\ell_1=0$. Unless \eqref{cond_sd} is violated, one describes a theory that is consistent with any length scales, provided its existence. In this sense, $U_0$ is the largest neighborhood of $\Gamma(\mathcal{C}_t)$ that complies with the criteria \eqref{cond_sd}.  

Suppose it is possible to contract the neighborhood $U_0$
of $\phi_0$
sufficiently to comply with \eqref{cond_sd}, then 
$\mathcal{A}_0$ can be described by an effective field theory 
with a short-distance cut-off $\ell$ beyond which a more fundamental theory is required.

This, however, does not imply that admissible fluctuations 
respect the background configuration. 
Relative to $\mathcal{A}_0$, the background seems stable 
provided that
\begin{eqnarray}
\label{cond_bs}
    \max_{\mathcal{O}\in\mathcal{A}_0}
    \Big(
        N_{\phi_0,\Psi_t}
        \big(
                \mathcal{R} (\mathcal{O})/
                \mathcal{R} (\mathcal{O}_0)
        \big)
    \Big)
    \leq
    \delta 
    \; ,
\end{eqnarray}
where  
$\delta$ is assumed to be in the open interval $(0,1)$. 
The value $\delta=0$ is excluded since 
it corresponds 
to a trivial fluctuation relative to $\mathcal{A}_0$,
and $\delta=1$ is excluded since it corresponds to
fluctuations that equalize the background observable
with respect to some quantum observable included 
in $\mathcal{A}_0$. Such fluctuations will trigger non-negligible backreactions on the classical observable. In this sense, $\delta$ has the meaning of a resolution scale
relative to $\mathcal{A}_0$ and quantifies to which extent the
background configuration can be distinguished from fluctuations
based on the local quantum observables in $\mathcal{A}_0$
if the system is in a state represented by the wave functional 
$\Psi_t$. It is possible to develop this interpretation 
further by considering actual measurement processes, but 
we leave this for future work. 
Note that only the concept of background stability relative 
to a chosen set of local quantum observables is sensible.
Relative background stability might imply a further 
contraction of $U_0$. 

The contractions of $U_0$ entailed by \eqref{cond_sd} and 
\eqref{cond_bs} guarantee a faithful description 
of those fluctuations respecting the background configuration 
within the framework of effective field theories 
at the level of local quantum observables contained in $\mathcal{A}_0$.
We will always assume that $U_0$ has been contracted
accordingly and denote the restriction of the formal measure 
space $(\Gamma(\mathcal{C}_t),\myd\phi)$ in accordance with 
the consistency conditions 
(\ref{cond_sd}) and (\ref{cond_bs}) 
by $(U_0,\myd\phi)$. Accordingly, $U_0$ marks the range of validity for the considered effective framework.

At last, we emphasize that our criterion in \eqref{cond_bs} is general:   the choice of $\mathcal O_0$ should not be seen as limited  to quantities derived solely from the classical field $\phi_0$ \footnote{
For example, in the case of a free quantum field, a comparison with classical observables based on classical fields is meaningless, as one can find a trivial background, i.e., $\phi_0 \equiv 0$.}. 
The comparison between  $\mathcal O$ and $\mathcal O_0$ is meaningful only 
 when it aligns with the specific questions being addressed.
Suppose we are concerned with the stability of the classical spacetime geometry. 
In this case, the appropriate quantities to compare will involve the expectation value of the Hamilton operator for a given quantum state, relative to geometric quantities that determine the energy scale of the curvature. Examples are the black hole mass in Schwarzschild spacetimes, or the square of the Hubble parameter in FLRW spacetimes.

\subsection{Analysis of Operators}
\label{sec:aoo}
In this section we prepare the spectral analysis \cite{ree80,ree75,ree79}
of various functional operators. This analysis 
requires to introduce smooth approximations of the identity 
in $\Gamma(\mathcal{C}_t)$ and is limited 
to an important subset of multilocal functions, given below, 
including Gaussian states. This limitation does not correspond to a mathematical 
requirement, it is rather a convenience for explicit computations. 

Let $U\subseteq U_0$ denote a subset of $\Gamma(\mathcal{C}_t)$ in accordance with 
the conditions \eqref{cond_sd} and 
\eqref{cond_bs}, consisting of fluctuations 
around a stable background in the domain of some effective field theory. 
The hypersurfaces $\Sigma_t$ 
parameterize the sections of the real line bundles $\mathcal{C}_t$
at any permissible time, and $U$ is bounded by two sections 
$\varphi_{_\mathrm{L}}$ and $\varphi_{_\mathrm{H}}$ in $\Gamma(\mathcal{C}_t)$ which
comprise its geometric boundary relative to $\mathbb{R}$ along $\Sigma_t$.
In other words, for $U=\cup_{p\in\Sigma_t}U_p$, $U_p=[\varphi_{_\mathrm{L}}(p),\varphi_{_\mathrm{H}}(p)]\subset\mathbb{R}$ 
at any point in the hypersurface $\Sigma_t$.
For better readability we refrain from specifying boundaries
by their respective base manifolds. 

We introduce the one parameter family 
$\{\mathcal{I}^\varepsilon\}_{\varepsilon\in (0,1)}$
of approximate identities
as follows: let $\mathcal{I}^\varepsilon: \Gamma(\mathcal{C}_t)$
$\rightarrow$ $\mathbb{R}^+$ with support contained in $U$
so that its integral over $U$ equals to one. 
On each fiber $\Gamma_p(\mathcal{C}_t)$, this family is given by the usual family 
$\delta_p^\varepsilon$ of Dirac distributions in $\Gamma_p^{\; *}(\mathcal{C}_t)$.
The relation between the global and local versions is
$\mathcal{I}^\varepsilon = \Pi_{p\in\Sigma_t} \delta^\varepsilon_p$.

In this article, we focus on a trivial but important 
subset of so-called multilocal functionals:
\begin{eqnarray}
\label{mlf}
    \mathcal{F}(\varphi) =  
    \sum_{n\in\mathbb{N}_0} \mathcal{K}^{(n)} \left(\varphi^{\otimes n}\right) \; ,
\end{eqnarray}
where $\mathcal{K}^{(n)}$ 
is a complex valued functional in $(\Gamma(\mathcal{C}_t)^{\otimes n})^*$, 
and it is understood that $\mathcal{K}^{(0)}$ is a complex number. 
This subset is trivial in the sense that each copy of the 
jet expansion of the configuration field terminates 
at the lowest order contribution. And it is important 
since it contains Gaussian states: Choose $\mathcal{K}^{(0)}=1$, 
$\kappa\in\mathbb{C}$
and for $n>0$,
\begin{eqnarray}
    \mathcal{K}^{(n)}(\varphi^{\otimes n})
    =
    \frac{\delta_{n,2m}}{m!}\left(\frac{\kappa}{2}\right)^m
    \left( \mathcal{K}^{(2)}\left(\varphi^{\otimes 2}\right)
    \right)^{m}
    \; .\nonumber
\end{eqnarray}
where $m,n \in \mathbb N$. 
Let $e_p^{\; *}$ be the evaluation functional returning 
the value of its argument at $p\in\Sigma_t$, 
and let $e_p$ denote its dual. We expand
\begin{eqnarray}
    \varphi
    =
    \int\limits_{p\in\Sigma_t} e_p^{\; *}(\varphi) \, e_p
    \; ,
\end{eqnarray}
and find by direct computation
\begin{eqnarray} \label{guassian_exp_functional}
    \mathcal{F}(\varphi)
    =
    \exp{
        \left(
            \frac{\kappa}{2}\; \mathcal{K}^{(2)}\left(\varphi^{\otimes 2}\right)
        \right)
        }
\end{eqnarray}
for a suitable configuration space. 

Let $\mathcal{F}$ be a multilocal functional of the form (\ref{mlf})
on $U$. 
For $\phi\in U$, consider the one parameter family of functionals 
\begin{eqnarray}
\label{Feps}
    \mathcal{F}^\varepsilon(\phi)
    =
    \int_U \myd\varphi \; \mathcal{I}^\varepsilon(\phi-\varphi) \mathcal{F}(\varphi)
    \; .
\end{eqnarray}
We compare both functionals $\mathcal{F}^\varepsilon$ and $\mathcal{F}$:
\begin{eqnarray}
   \left|\mathcal{F}^\varepsilon(\phi) - \mathcal{F}(\phi)\right|
   &\leq&
   \int_{U}\myd\varphi \; 
   \mathcal{I}^\varepsilon(\phi-\varphi) 
   \left|\mathcal{F}(\varphi) - \mathcal{F}(\phi)\right|
   \nonumber \\
   &\leq&
   \sup_{S_U} \left|\mathcal{F}(\varphi) - \mathcal{F}(\phi)\right|
    \; , \nonumber
\end{eqnarray}
where the supremum is taken over 
$S_U:=\{\varphi\in U: \sup_{\Sigma_t} |\varphi-\phi|<\varepsilon\}$.
The first estimate follows from the triangle inequality
and the properties of the approximate identity, and
the second from the definition of the supremum as the 
least upper bound estimate. Again, this inequality uses properties of the approximate identity. 
Whenever $\varepsilon\rightarrow 0^+$, $\varphi$
is required to approximate $\phi$ to an ever better degree. 
As a consequence, the right hand side of the above inequality 
converges to zero and, therefore, $\mathcal{F}^\varepsilon\rightarrow \mathcal{F}$
in $L^2(U,\myd\varphi)$.

In the remainder of this section, we 
highlight informally some construction of
functional differential operators. At this stage, the discussion is kept 
at a formal level without considering 
questions pertinent to infinite dimensional analysis including, 
in particular, domain questions. These will be settled 
in Sec. \ref{sec:fmo} that is devoted to a thorough 
analysis of the functional momentum operator and its spectral properties. 

Let $\mathcal{K}^{(n)}\in (\Gamma_\mathrm{c}(\mathcal{C}_t)^{\otimes n})^*$, the functional differential $\delta$ is characterized by
\begin{eqnarray}
    &&\delta:
    (\Gamma_\mathrm{c}(\mathcal{C}_t)^{\otimes n})^*
    \rightarrow 
    (\Gamma_\mathrm{c}(\mathcal{C}_t)^{\otimes (n+1)})^*,
    \nonumber \\
    && \delta \mathcal{K}^{(n)}
    \left(\varphi^{\otimes n}\otimes f\right)
    :=
    \left(f,\delta_\varphi\right) \mathcal{K}^{(n)}
    \left(\varphi^{\otimes n}\right)\, ,
\end{eqnarray}
for all $n\in\mathbb{N}$, 
which is a convenient 
way to store all the functional derivatives of multilocal functionals. 
In the exceptional case of a local functional, the functional differential 
is given by exchanging the argument of the functional with the direction
of the functional derivative, 
$
    \delta \mathcal{K}^{(1)}(\varphi\otimes f)
    = \mathcal{K}^{(1)}(f)
$.

The functional gradient $\mathrm{Grad} \, \mathcal{K}^{(n)}$
of a multilocal functional $\mathcal{K}^{(n)}$ is the multilocal functional 
$\delta \mathcal{K}^{(n)}(\varphi^{\otimes n}, \cdot)$. 
In DeWitt's condensed notation \cite{dewitt2003global}, 
\begin{eqnarray}\label{funcGrad}
    \mathrm{Grad} \, \mathcal{K}^{(n)} 
    =
    \delta_{\varphi(p)}\;  \mathcal{K}^{(n)}
    \otimes  e_p^{\; *}
    \; .
\end{eqnarray} 
The functional Hessian $\mathrm{Hess} \, \mathcal{K}^{(n)}$ is
its second functional differential $\delta(\delta \mathcal{K}^{(n)})$.
In greater detail, 
\begin{eqnarray}
    \mathrm{Hess} \, \mathcal{K}^{(n)}
    =
    \delta_{\varphi(p_1)}\delta_{\varphi(p_2)} \, \mathcal{K}^{(n)}
    \otimes e_{p_1}^{\; *} \otimes e_{p_2}^{\; *} 
    \; .
\end{eqnarray}
The functional Laplacian $\mathrm{Div}(\mathrm{Grad} \, \mathcal{K}^{(n)})$
is the contraction 
$C_{n+1,n+2}(\mathrm{Hess}\, \mathcal{K}^{(n)})$ which for a bi-local functional yields
$C_{1,2}(K(p_1,p_2) e_{p_1}^{\; *}\otimes e_{p_2}^{\; *})=K(p,p)$
for some function $K:\Sigma_t\times\Sigma_t\rightarrow\mathbb{C}$.

\subsection{Functional momentum operator} \label{sec:fmo}
In this section we present the spectral analysis of the 
functional momentum operator $\mathcal{P}$ on domains subject to 
boundaries in configuration field space.

We proceed in three steps. The first is of pedagogical nature 
and prepares us for the proof that $\mathcal{P}^\mathrm{cl}$ extends 
$\mathcal{P}_1$, where $\mathcal{P}^\mathrm{cl}$ denotes the closure 
of $\mathcal{P}$ with domain 
$\mathcal{C}_U^\infty (\Gamma_\mathrm{c}(\mathcal{C}_t),\mathbb{C})$
and $\mathcal{P}_1$ is defined on the domain 
$\mathcal{C}_U^1 (\Gamma_\mathrm{c}(\mathcal{C}_t),\mathbb{C})$.
Again, $U$ denotes the compactum in the space of 
instantaneous field configuration in accordance with
\eqref{cond_sd} and 
\eqref{cond_bs}. As a consequence, these domains require to specify conditions on the boundary $\partial U$.

In the next step, we show 
that $\mathcal{P}$ is symmetric on a certain pre-Hilbert space 
and determine its adjoint on a related, but less restrictive, pre-Hilbert space. 
It remains to argue that these pre-Hilbert spaces are, in fact, complete 
with respect to the norm induced by the inner product.
Finally, we derive that $\mathcal{P}$ is not essentially self-adjoint 
and classify all self-adjoint extensions. Since the functional momentum operator 
is so simple, this can be done by direct computation instead
of employing von Neumann's deficiency indices \cite{ree80}. 

Without going right away into details about the domain of the
functional momentum operator, it certainly encompasses 
$\mathcal{C}_U^\infty (\Gamma_\mathrm{c}(\mathcal{C}_t),\mathbb{C})$. 
So for the moment, let 
$\mathrm{Dom}(\mathcal{P})=
\mathcal{C}_U^\infty (\Gamma_\mathrm{c}(\mathcal{C}_t),\mathbb{C})$
and $\mathrm{Dom}(\mathcal{P}_1)=\mathcal{C}_U^1 (\Gamma_\mathrm{c}(\mathcal{C}_t),\mathbb{C})$, be subjected to boundary conditions on $\partial U$.
It is useful to work with the functional generalization of
directional derivatives: for $f\in \mathcal{C}^\infty_\mathrm{c}(\Sigma_t)$, 
let
\begin{eqnarray}
    \mathcal{P}(f) \; \Psi
    := 
    \int\limits_{p\in\Sigma_t} f(p) \mathcal{P}_{\hspace{-0.1cm}\varphi(p)}
    \; \Psi(\varphi) \equiv (f,\mathcal{P}_{\varphi}) \Psi(\varphi)
    \; ,
\end{eqnarray}
where $\mathcal{P}_{\hspace{-0.1cm}\varphi(p)} = -\mathrm{i} \delta_{\varphi(p)}$
denotes the functional derivative in the tangent space to the 
fiber located over $p\in\Sigma_t$ in field configuration 
space, if $\Psi\in \mathrm{Dom}(\mathcal{P})$, and accordingly 
for $\mathcal{P}_1$. Note that the functional derivative 
includes a factor $1/\sqrt{\mathrm{det}(q)}$ evaluated at $p$.
Since the induced volume factor is smooth, it can be absorbed 
in $f$.

The approximate identity $\mathcal{I}^\varepsilon$ constructed above is by assumption in $\mathrm{Dom}(\mathcal{P})$, 
so we can evaluate $\mathcal{P}$ on the one-parameter family of
functionals $\Psi^\varepsilon$: 
\begin{eqnarray}
    \mathcal{P}(f) \Psi^\varepsilon (\phi)
    &=& 
    \int\limits_U \myd\varphi \; (f,\mathcal{P}_\phi) \mathcal{I}^\varepsilon
    (\phi-\varphi) \Psi(\varphi)
   \nonumber \\
    &=&
    \int\limits_U \myd\varphi \; (-f,\mathcal{P}_\varphi) \mathcal{I}^\varepsilon
    (\phi-\varphi) \Psi(\varphi)
    \nonumber \\
    &=&
    \int\limits_U \myd\varphi \; \mathcal{I}^\varepsilon(\phi-\varphi)
    \mathcal{P}_1(f) \Psi(\varphi) 
    +
    \mathcal{B}^\varepsilon_{\partial U}(\Psi,f)
    \; .\nonumber
\end{eqnarray}
In order to comply with 
conditions (\ref{cond_sd}) and 
(\ref{cond_bs}), the boundary term $\mathcal{B}^\varepsilon_{\partial U}(\Psi,f)$
is required to vanish, 
which restricts the domain of $\mathcal{P}$ further. 
Explicitly, 
\begin{eqnarray}
   \mathcal{B}^\varepsilon_{\partial U}(\Psi,f)
   &=&
   -\mathrm{i} \hspace{-0.1cm}
   \int\limits_{p\in\Sigma_t} \hspace{-0.1cm}f(p) \hspace{-0.1cm}
   \int\limits_{U\setminus U_p} \hspace{-0.1cm}\myd\varphi\; 
   \left[\mathcal{I}^\varepsilon(\phi-\varphi) \Psi(\varphi)\right]_{\partial U_p} = 0\;  \nonumber
\end{eqnarray} 
Our previous discussion following \eqref{Feps} demonstrated that 
since $\Psi^\varepsilon$ has support on
a fixed compact subset in the space of instantaneous field configurations, 
$\Psi^\varepsilon\rightarrow \Psi$ in $L^2(U,\myd\varphi)$, and now
we find similarly that  
$\mathcal{P}(f) \Psi^\varepsilon (\phi) \rightarrow \mathcal{P}_1(f) \Psi(\phi)$
in $L^2(U,\myd\varphi)$, as $\varepsilon\rightarrow 0^+$.
This shows that the closure of $\mathcal{P}$ contains $\mathcal{P}_1$.

To determine the adjoint $\mathcal{P}^*$ of $\mathcal{P}$, 
we introduce below the notion of absolute continuous functionals. 
Consider the formal measure space $L^2(U,\myd\varphi)$ and 
for each $\delta>0$, define $\mathcal{S}_\delta$ to be the collection 
of all countable families of pairwise disjoint open connected sets 
$\{E_a\}_{a\in\mathbb{N}}$ with $\sum_{a\in\mathbb{N}} \mu(E_a)<\delta$, 
where $\mu(E_a)$ denotes the finite volume of $E_a\subset U$.  
We call a functional 
$\Psi:U\subset \Gamma^\infty_\mathrm{c}(\mathcal{C}_t)\rightarrow\mathbb{C}$
absolutely continuous if for every $\varepsilon>0$ there is a $\delta>0$
so that for any $\{D_a\}_{a\in\mathbb{N}}\in \mathcal{S}_\delta$ 
we have 
\begin{eqnarray}
    \sum\limits_{a\in\mathbb{N}} \,
    \sup
    \left(
        |\Psi(\varphi_1)-\Psi(\varphi_2)|_\mathbb{C} 
        : \varphi_1 \, , \varphi_2 \in D_a
    \right) < \varepsilon
    \; .
\end{eqnarray}
We denote the set of all absolutely continuous functionals 
on $U$ with values in $\mathbb{C}$ by $\mathcal{AC}(U)$.
The subset of all bounded functionals in $\mathcal{AC}(U)$
is an algebra over $\mathbb{C}$.

Let 
$
    \mathrm{Dom}(\mathcal{P})
    =
    \{\Theta\in\mathcal{AC}(U): \Theta|_{\partial U}=0\}
$
and, as before, $\mathcal{P}(f)=(f,-i\delta_\varphi)$.
The functional momentum operator is densely defined and 
an integration by parts shows that it is symmetric. 
Let $\Psi\in\mathrm{Dom}(\mathcal{P}^*)$.
Then, by definition, the adjoint satisfies \cite{ree80,ree75}
\begin{eqnarray}
\label{adjoint}
    \left\langle
        \mathcal{P}(f) \mathcal{X}^\varepsilon_{U_1},\Psi
    \right\rangle
    =
    \left\langle
        \mathcal{X}^\varepsilon_{U_1}, \mathcal{P}(f)^*\Psi
    \right\rangle
    \; ,
\end{eqnarray}
where $U_1\subset U$ and $\mathcal{X}^\varepsilon_{U_1}$
is a smooth approximation of the indicator functional, which is locally 
characterized by 
$\mathcal{X}^\varepsilon_{U_{1p}}(\varphi(p))=1$ if 
$\varphi(p)\in U_{1p}\subset\mathbb{R}$ for every 
$p\in\Sigma_t$, and zero otherwise.
The right hand side of (\ref{adjoint}) is readily computed:
\begin{eqnarray}
    \left\langle
        \mathcal{X}^\varepsilon_{U_1}, \mathcal{P}(f)^*\Psi
    \right\rangle
    \stackrel{\varepsilon\rightarrow 0^+}{\longrightarrow}
    \int\limits_{U_1} \myd\varphi \; \mathcal{P}(f)^*\Psi
    \; .
\end{eqnarray}
The left hand side of (\ref{adjoint}) requires more work.
\begin{eqnarray}
    \left\langle
        \mathcal{P}(f) \mathcal{X}^\varepsilon_{U_1},\Psi
    \right\rangle
    &=&
    \int\limits_{U} \myd\varphi\; 
     (f,\mathrm{i}\delta_\varphi)
     \left(\mathcal{X}^\varepsilon_{U_1} \Psi\right)(\varphi)
    \nonumber \\
    &&+
    \int\limits_{U} \myd\varphi\; \mathcal{X}^\varepsilon_{U_1}(\varphi)
    (f,-\mathrm{i}\delta_\varphi) \Psi (\varphi)
    \; .
\end{eqnarray}
The first term on the right hand side of this equation vanishes since 
$\partial U \cap U_1=\emptyset$ by construction.
In the limit $\varepsilon\rightarrow 0^+$ equation (\ref{adjoint})
becomes 
\begin{eqnarray}
    \int\limits_{U_1} \myd\varphi \; \mathcal{P}(f)^*\Psi(\varphi)
    =
    \int\limits_{U_1} \myd\varphi \; (f,-\mathrm{i}\delta_\varphi)\Psi(\varphi)
    \; .
\end{eqnarray}
So fiberwise we have 
\begin{eqnarray}
    -\mathrm{i}\hspace{-0.2cm}
    \int\limits_{U_1\setminus U_{1p}} \myd\varphi \; \Psi(\varphi)|_{\partial U_{1p}}
    =
    \int\limits_{U_1} \myd\varphi \; \mathcal{P}_{\varphi(p)}^*\Psi(\varphi)
    \; ,
\end{eqnarray}
which means $\Psi$ is absolutely continuous, $\Psi\in\mathcal{AC}(U)$, and 
$\mathcal{P}(f)^*\Psi(\varphi) = (f,-\mathrm{i}\delta_\varphi)\Psi(\varphi)$.
The other direction is similar: integration by parts shows that any 
$\Psi\in\mathcal{AC}(U)$ is in the domain of $\mathcal{P}^*$
and $\mathcal{P}^*\Psi = (f,-\mathrm{i}\delta_\varphi)\Psi$ for 
$\Psi\in \mathrm{Dom}(\mathcal{P}^*)$.
Therefore $\mathrm{Dom}(\mathcal{P}^*)=\mathcal{AC}(U)$
and $\mathcal{P}^*=-\mathrm{i} \delta_\varphi$.

Clearly, $\mathrm{Dom}(\mathcal{P}^*) \supset \mathrm{Dom}(\mathcal{P})$.
Moreover, $\mathcal{P}$ is not essentially self-adjoint. 
Just consider \cite{Schneider:2018tyr}
\begin{eqnarray}
    \Theta(\varphi) := \exp\left(\pm \mathcal{K}^{(1)}(\varphi)\right)
    \in \mathrm{Dom}(\mathcal{P}^*) \; ,
\end{eqnarray}
where $\mathcal{K}^{(1)}$ is a real valued functional on the 
compactly supported smooth sections of the configuration bundle
$\mathcal{C}_t$,
and $\mathcal{P}(f)\Theta(\varphi)=\mp \mathrm{i} \, \mathcal{K}^{(1)}(f)\Theta(\varphi)$.

This raises the question whether $\mathcal{P}$ has any self-adjoint 
extensions. The answer, given below, is positive and constructive
since it provides a complete classification of all self-adjoint extensions.
Let $\mathcal{P}(f)$ be the operator $(f,-\mathrm{i}\delta_\varphi)$
on $L^2(U,\myd\varphi)$ with the domain 
$\mathrm{Dom}(\mathcal{P})=\{\Psi\in\mathcal{AC}(U): \Psi|_{\partial U}=0\}$. 
We showed above that $\mathcal{P}$ is symmetric on this domain and that
the adjoint $\mathcal{P}^*$
is the operator $(f,-\mathrm{i}\delta_\varphi)$
with domain $\mathcal{AC}(U)$.
For $\Theta\in\mathrm{Dom}(\mathcal{P})$ and $\Psi\in\mathrm{Dom}(\mathcal{P}^*)$
consider the difference
\begin{eqnarray}
    \Delta( \mathcal P(f))
    :=
    \left\langle
        \mathcal{P}(f)\Theta,\Psi
    \right\rangle
    -
    \left\langle
        \Theta,\mathcal{P}(f)^*\Psi
    \right\rangle
    \; .
\end{eqnarray}
Since $\mathrm{Dom}(\mathcal{P})\subset\mathrm{Dom}(\mathcal{P}^*)$
we perform an integration by parts in the second term and find 
\begin{eqnarray}
\label{saext}
    \Delta( \mathcal P(f))
    =
    \mathrm{i}
    \int\limits_{p\in\Sigma_t} f(p)
    \left\langle
        \Theta, \Psi
    \right\rangle|_{\partial U_{p}}
    \; ,
\end{eqnarray}
where 
$
    \left\langle\Theta, \Psi\right\rangle|_{\partial U_{p}}
    :=
    \left\langle\Theta, 
    \mathcal{X}_{\partial U_p}
    \Psi\right\rangle
$ 
with the following sign convention implied: 
$
\mathcal{X}_{\partial U_p}
=
\, ^\mathrm{D}\hspace{-0.04cm}\delta_{\varphi_\mathrm{H}(p)} \circ e_p^{\; *}
- \, ^\mathrm{D}\hspace{-0.04cm}\delta_{\varphi_\mathrm{L}(p)}\circ e_p^{\; *}
$.
For a self-adjoint extension $\Delta(\mathcal{P}(f))$ is necessarily required to be zero, 
which is precisely enforced by the boundary conditions 
specified in the domain of the functional momentum operator. Note that this holds 
without imposing any boundary conditions on those wave functionals 
in the domain of the corresponding adjoint operator. Therefore, the functional 
momentum operator defined this way is not self-adjoint. 

Assume there exists a self-adjoint extension $\Pi$ of $\mathcal{P}$
and choose a $\Theta\in \mathrm{Dom}(\Pi)\setminus\mathrm{Dom}(\mathcal{P})$.
Then \eqref{saext} demands that 
\begin{eqnarray} \label{SA:condition_0}
    \int\limits_{p\in\Sigma_t} f(p) \left\|\Theta\right\|^2_{L^2}|_{\partial U_p}
    =
    0
    \; ,
\end{eqnarray}
where 
$
    \|\Theta\|^2_{L^2}|_{\partial U_p}
    :=
    \langle \Theta, \mathcal{X}_{\partial U_p}\Theta\rangle
$.
For $\mathrm{A}\in\{\mathrm{L},\mathrm{H}\}$, let 
\begin{eqnarray}
    \zeta_{_\mathrm{A}}(p) 
    :=
    \varphi_{_\mathrm{A}}(p) \, e_p + \int\limits_{p^\prime} \left(\mathcal{X}_{\Sigma_t\setminus\{p\}}\varphi\right)(p^\prime) \, e_{p^\prime}
    \; .
\end{eqnarray}
Since by assumption $\Theta\notin \mathrm{Dom}(\mathcal{P})$, equation
(\ref{saext}) can become zero if and only if there exists
a complex valued function $\alpha$ on $\Sigma_t$ with unit modulus so that 
\begin{eqnarray} \label{SA_condition}
    \Theta(\zeta_{_\mathrm{H}}(p)) - \alpha(p) \Theta(\zeta_{_\mathrm{L}}(p))
    = 
    0
\end{eqnarray}
is true at any point on the hypersurface $\Sigma_t$. 
This condition is required to hold for any functional in $\mathrm{Dom}(\Pi)$
with the same function $\alpha$.
Introduce $\mathcal{P}_\alpha(f) = (f,-\mathrm{i}\delta_\varphi)$ on 
\begin{eqnarray}
    \left\{
        \Psi\in\mathcal{AC}(U): 
        \Psi(\zeta_{_\mathrm{H}}(p)) 
        = \alpha(p) \Psi(\zeta_{_\mathrm{L}}(p)) \; \forall p\in\Sigma_t
    \right\} \; .
\end{eqnarray}
Clearly $\mbox{Dom}(\Pi)\subset \mbox{Dom}(\mathcal{P}_\alpha)$. 
Since $\mathcal{P}_\alpha$ is symmetric and $\Pi$ is by assumption 
a self-adjoint extension of $\mathcal{P}$, we have $\Pi=\mathcal{P}_\alpha$
for some function $\alpha:\Sigma_t\rightarrow\mathbb{C}$ with 
$|\alpha|_\mathbb{C}=1$.

It remains to determine which of the symmetric extensions $\mathcal{P}_\alpha$ 
are in fact self-adjoint. Let $\Theta\in\mathrm{Dom}(\mathcal{P}_\alpha)$
and $\Psi\in \mathrm{Dom}(\mathcal{P}_\alpha^{\, *})$.
Then \eqref{saext} being zero requires 
\begin{eqnarray}
    \alpha^*(p) \Theta^*(\zeta_{_\mathrm{L}}(p))
    \Psi(\zeta_{_\mathrm{H}}(p))
    =
   \Theta^*(\zeta_{_\mathrm{L}}(p))\Psi(\zeta_{_\mathrm{L}}(p)) 
\end{eqnarray}
at every point $p$ in the hypersurface $\Sigma_t$, which locally has 
the following solution: 
$\Psi(\zeta_{_\mathrm{H}}(p)) = \alpha(p) \Psi(\zeta_{_\mathrm{L}}(p))$.
But then $\Psi\in \mathrm{Dom}(\mathcal{P}_\alpha)$, and we are safe to 
conclude 
$\mathrm{Dom}(\mathcal{P}_\alpha^{\, *})=\mathrm{Dom}(\mathcal{P}_\alpha)$. 
In other words, all symmetric extensions $\mathcal{P}_\alpha$, 
indexed by a complex valued function on $\Sigma_t$ with unit modulus, 
are self-adjoint. 
Apparently we have found infinitely many self-adjoint extensions which describe different physics \cite{ree75}.
Of course, this is to be expected from the point particle limit. 

In the following section, we will extend our analysis to find the self-adjoint domain of the Hamilton operator. This will bring us one step closer to studying the consequences of the dynamics it generates.

\subsection{Hamilton Operator} \label{sec:ho}
As a proof of concepts, it is sufficient to consider the Hamiltonian of a free field. 
The operator of interest is therefore the functional generalization 
of the Laplacian introduced above at the informal level. 
Discussing the spectral properties of the functional 
Laplacian is an application of quadratic form techniques 
and positivity based on a theorem by von Neumann 
which was originally proven \cite{ree75} using operator-theoretic techniques 
in the case of point particles.

Let 
$
    \mathcal{Q}: \mathrm{Dom}(\mathcal{P})\times\mathrm{Dom}(\mathcal{P})
    \rightarrow \mathbb{C}
$
be a quadratic form on the form domain $\mathrm{Dom}(\mathcal{P})$, 
defined by $\mathcal{Q}(\Theta,\Psi):=\langle \mathcal{P}\Theta,\mathcal{P}\Psi\rangle$, 
so that
$\mathcal{Q}(\cdot,\Psi)$ is conjugate linear and 
$\mathcal{Q}(\Theta,\cdot)$ is linear for $\Theta,\Psi\in \mathrm{Dom}(\mathcal{P})$. 
Apparently, $\mathcal{Q}$ is nonnegative and since $\mathcal{P}$ is closed 
as an operator, $\mathcal{Q}$ is closed as a quadratic form.
Then there is a unique self-adjoint operator $\mathcal{T}_\varphi$
associated with $\mathcal{Q}$. This follows from the Riesz lemma, 
the Hellinger-Toeplitz theorem, and by direct application of 
the spectral theorem in multiplication operator form.
We will now determine the operator $\mathcal{T}_\varphi$ that is associated with the kinetic part of the Hamilton operator. 

Note that $\mathrm{Dom}(\mathcal{P})$ is a subspace of 
the Hilbert space $L^2(\Gamma(\mathcal{C}_t),\myd\varphi)$
under the inner product 
$
    \langle\langle\Theta,\Psi\rangle\rangle
    :=
    \mathcal{Q}(\Theta,\Psi)+\langle\Theta,\Psi\rangle
$.
Denote by $(\mathrm{Dom}(\mathcal{P}))^*$ the space of 
bounded conjugate linear functionals on $\mathrm{Dom}(\mathcal{P})$.
Introduce the linear embedding
of $L^2(\Gamma(\mathcal{C}_t),\myd\varphi)$ into $(\mathrm{Dom}(\mathcal{P}))^*$ 
by $\Psi\mapsto \langle\cdot, \Psi\rangle$. 
From the Cauchy-Schwarz inequality it follows that this map is bounded. 
Since the identity map embeds $\mathrm{Dom}(\mathcal{P})$
in $L^2(\Gamma(\mathcal{C}_t),\myd\varphi)$, we have 
the following chain of inclusions: 
$
    \mathrm{Dom}(\mathcal{P})
    \subset
    L^2(\Gamma(\mathcal{C}_t),\myd\varphi)
    \subset
    (\mathrm{Dom}(\mathcal{P}))^*.
$ Define 
$
    \bar{\mathcal{P}}^*:L^2(\Gamma(\mathcal{C}_t),\myd\varphi)
    \rightarrow 
    (\mathrm{Dom}(\mathcal{P}))^*
$
by $(\bar{\mathcal{P}}^*\Theta)(\Psi):=\langle \Theta,\mathcal{P}\Psi\rangle$.
From the definition of the formal adjoint we see that 
$
    \mathrm{Dom}(\mathcal{P}^*)
    =
    \{\Theta: \bar{\mathcal{P}}^*\Theta \in L^2(\Gamma(\mathcal{C}_t),\myd\varphi)\}
$
and $\mathcal{P}^*$ equals $\bar{\mathcal{P}}^*$ restricted to 
$\mathrm{Dom}(\mathcal{P}^*)$.
Next, let 
$
    \bar{\mathcal{T}}_\varphi: 
    \mathrm{Dom}(\mathcal{P})
    \rightarrow
    (\mathrm{Dom}(\mathcal{P}))^*
$ 
be the map given by
$\langle\bar{\mathcal{T}}_\varphi\Theta,\Psi\rangle:=\mathcal{Q}(\Theta,\Psi)$.
The proof of the theorem associating a (unique) self-adjoint 
operator to a closed positive quadratic form mentioned above 
gives 
$
    \mathrm{Dom}(\mathcal{T}_\varphi)
    =
    \{
        \Theta\in \mathrm{Dom}(\mathcal{P}): 
        \bar{\mathcal{T}}_\varphi\Theta
        \in
        L^2(\Gamma(\mathcal{C}_t),\myd\varphi)
    \}
$
and $\mathcal{T}_\varphi$ equals $\bar{\mathcal{T}}_\varphi$ restricted 
to $\mathrm{Dom}(\mathcal{T}_\varphi)$.
Let $\Theta,\Psi\in\mathrm{Dom}(\mathcal{P})$.
Then 
$
    (\bar{\mathcal{P}}^*(\mathcal{P}\Theta))(\Psi)
    =
    \mathcal{Q}(\Theta,\Psi)
    =
    (\bar{\mathcal{T}}_\varphi\Theta)(\Psi)
$. Hence, $\bar{\mathcal{T}}_\varphi=\bar{\mathcal{P}}^*\mathcal{P}$.
Therefore, 
\begin{eqnarray}\label{funcLap}
    \mathrm{Dom}(\mathcal{T}_\varphi)
    &=&
    \{
        \Theta\in \mathrm{Dom}(\mathcal{P}): 
        \bar{\mathcal{T}}_\varphi\Theta
        \in
        L^2(\Gamma(\mathcal{C}_t),\myd\varphi)
    \}
    \nonumber \\
    &=&
   \{
        \Theta\in \mathrm{Dom}(\mathcal{P}): 
        \bar{\mathcal{P}}^*(\mathcal{P}\Theta)
        \in
        L^2(\Gamma(\mathcal{C}_t),\myd\varphi)
    \} 
    \nonumber \\
    &=&
    \{
        \Theta\in \mathrm{Dom}(\mathcal{P}): 
        \mathcal{P}\Theta\in\mathrm{Dom}(\mathcal{P}^*)
    \}
    \nonumber \\
    &=&
    \mathrm{Dom}(\mathcal{P}^*\mathcal{P})
    \; ,
\end{eqnarray}
where the first equality is just the quoted result 
on the domain of $\mathcal{T_\varphi}$, the second equality 
follows from $\bar{\mathcal{T}}_\varphi=\bar{\mathcal{P}}^*\mathcal{P}$,
and the third and fourth are straightforward. 
Moreover, $\mathcal{T}_\varphi$ equals $\bar{\mathcal{T}}_\varphi$
restricted to the domain of $\mathcal{T}_\varphi$ which equals 
$\mathcal{P}^*\mathcal{P}$.
In other words, the self-adjoint operator associated with 
the closed positive quadratic form $\mathcal{Q}$ is
$\mathcal{P}^*\mathcal{P}$. 

Hence, we are ultimately equipped to determine the solutions to the functional Schr\"odinger equation for a free quantum field theory which we will then compare with the self-adjoint domain.

\section{ Dynamical aspects }\label{sec:da}

Our previous discussion showed that in a configuration field space subjected to boundaries at a finite distance, it is always possible to construct a momentum operator $\mathcal P $ and its quadratic form $\mathcal P^* \mathcal P$ with infinitely many self-adjoint extensions on a kinematic level. 
This mere possibility does not necessarily imply compatibility with the dynamics imposed by the Schr\"odinger equation for any $\Psi_t(\varphi)\in \text{Dom}(\mathcal P)$.

In this section, we present the dynamic analysis to accompany the kinematic statements.
We proceed in three steps: first, we investigate, for Gaussian wave-functionals, under which criterion the Hamilton operator $\mathcal{H}$ admits a self-adjoint extension on the kinematic level.
Second, we introduce the dynamics provided by the functional Schr\"odinger equation and examine the corresponding solution space $\Gamma_{\rm can}$. 
Third, we show that only static space-times provide the possibility for such self-adjoint extensions that are respected by the dynamics.
 
We begin with determining the self-adjoint domain $\mathrm{Dom}(\mathcal{P})$ of a Gaussian state $\Psi_t$ in $L^2(\Gamma(\mathcal{C}_t,\myd\phi)$. 
Consider
$
     \mathcal{AC}(U)\ni \Psi_t (\varphi) =\mathcal{N}_t \mathcal F_t(\varphi)
$
 where  $\mathcal F_t(\varphi)$ is given by \eqref{guassian_exp_functional} and  $\mathcal{N}_t\in\mathbb{C}$ denotes the normalization we will determine later. Then, $\eqref{SA:condition_0}$  demands
 \begin{align} \label{SA_condition:P}
    \int_{U\setminus U_p} \myd\varphi \,  \lvert  \mathcal F_t(\zeta_{_\mathrm{H}}(p)) \rvert^2 = \int_{U\setminus U_p} \myd\varphi \,  \lvert \mathcal F_t(\zeta_{_\mathrm{L}}(p)) \rvert^2 
\end{align}
where $\mathcal F(\zeta_{_\mathrm{A}}(p)) = \mathcal F(\varphi)\lvert_{\partial U_p}$, means that $\mathcal F(\varphi)$ is evaluated at the boundary $\partial U_p$, which sets  $\varphi = \zeta_{_\mathrm{A}}(p)$. 
The functional integral is given by  a consistency requirement that $\varphi_{_\mathrm{A}}(p)$ cannot be a functional of $\varphi(p')$ for all $p'\in \Sigma_t\setminus  \{p\} \equiv \Sigma_t^p$. 
Note that, while the condition \eqref{SA_condition:P} holds true when $\varphi_{_\mathrm{H}}(p)  = \varphi_{_\mathrm{L}}(p)$ for all $p\in\Sigma_t$, this only reflects the trivial choice of a set containing only a single point, i.e. $U=\{0\}$. 

Suppose $\varphi_{_\Lambda}(p) \equiv \varphi_{_\mathrm{H}}(p)  = -\varphi_{_\mathrm{L}}(p)$ for all $p\in \Sigma_t$. Then, the functional integral remains parity even, i.e. for all $p'\in \Sigma_t^p$ the map $\varphi(p') \rightarrow -\varphi(p')$ will not modify the integral in \eqref{SA_condition:P}. 
The important difference, however, is located in $\mathcal F_t(\zeta_{_\mathrm{A}}(p))$. Since 
\begin{align} 
    \zeta_{_\mathrm{A}}^{\otimes 2} &= \varphi_{_\mathrm{A}}(p)^2 e_p\otimes e_p \nonumber\\
    & \phantom{=}+ 2 \varphi_{_\mathrm{A}}(p) \int_{p'} (\mathcal X_{\Sigma_t^{p}} f\varphi)(p') e_p\otimes e_{p'} 
    \\ 
    & \phantom{=} + \int_{p',q'} (\mathcal X_{\Sigma_t^{p}} \varphi)(p') ( \mathcal X_{\Sigma_t^{p}} \varphi)(q') \,  e_{p'}\otimes e_{q'} \nonumber,
\end{align}
then for all $p' \in \Sigma_t^p$, sending $\varphi(p') \rightarrow -\varphi(p') $ yields
\begin{align}
     \int_{U\setminus U_p} \myd\varphi \,  \lvert  \mathcal F_t(\zeta_{_\mathrm{H}}) \rvert^2 \rightarrow \int_{U\setminus U_p} \myd\varphi \,  \lvert \mathcal F_t(\zeta_{_\mathrm{L}}) \rvert^2
\end{align}
which concludes that $\eqref{SA_condition}$ and $\eqref{SA_condition:P}$ can be satisfied, provided that $\varphi_{_\mathrm{H}}(p) = -\varphi_{_\mathrm{L}}(p)$ for all $p\in\Sigma_t$. 
Accordingly, there exist infinitely many self-adjoint extensions $\mathcal{P}_\alpha$ of $\mathcal P$ provided that
\begin{align}\label{domp}
    \text{Dom}(\mathcal P) = \{ &\Psi \in \mathcal{AC}(U) : \Psi(\zeta_{_\mathrm{H}}(p)) = \alpha(p) \Psi(\zeta_{_\mathrm{L}}(p)) , 
    \nonumber
    \\ &\phantom{=}\mbox{with}\;\; \varphi_{_\mathrm{H}}(p) = -\varphi_{_\mathrm{L}}(p) 
    \,  \forall p \in \Sigma_t \}.
\end{align}
An immediate follow-up question would be whether or not the Hamilton operator for a free scalar field admits similar extensions.
The answer to this question is in the positive and is mainly concerned with the spectrum of the kinetic operator $\mathcal P^* \mathcal P$. For $\mathcal P^* \mathcal P$ to be self-adjoint, $\mathcal P_{\varphi(p)} \Psi_t \equiv \Psi_t^{\prime} \in \text{Dom}(\mathcal P)$ must hold for any $p\in\Sigma_t$ where $\alpha(p)$ admits a unit modulus. 
This results in the following two relations coming from $\eqref{SA:condition_0}$ and \eqref{SA_condition} 
\begin{align}
    \Psi_t^{\prime}(\zeta_{_\mathrm{H}}(p))-\alpha(p) \Psi_t^{\prime}(\zeta_{_\mathrm{L}}(p))   = 0 , 
    \label{P2:SA1}
    \\ 
    \Psi_t^{\prime}(\zeta_{_\mathrm{H}}(p))-\alpha^*(p) \Psi_t^{\prime}{}^*(\zeta_{_\mathrm{L}}(p))    = 0 .
    \label{P2:SA2}
\end{align}
which provide two conditions under which $\mathcal P^*\mathcal P$ admits self-adjoint extensions. 
Since $\Psi$ is Gaussian, \eqref{P2:SA1} can be re-expressed as
\begin{align}
     \mathcal K ^{(2)}{}'&(\zeta_{_\mathrm{H}}(p)) \, \Psi_t(\zeta_{_\mathrm{H}}(p))
     - \alpha(p)    \mathcal K ^{(2)}{}'(\zeta_{_\mathrm{L}}(p))  \, \Psi_t(\zeta_\mathrm{L}(p))=0
    \nonumber  
\end{align}
where $\mathcal K^{(2)}{}'(\zeta_{_\mathrm{A}}(p)) \equiv \delta_{\varphi(p)} \mathcal K^{(2)}(\varphi) \lvert_{\varphi 
= \zeta_{_\mathrm{A}}(p)} $.
Using \eqref{SA_condition}, the above equation reduces to 
\begin{align} \label{P2:SA_condition_1}
    \Big( \mathcal K^{(2)}{}'(\zeta_{_\mathrm{H}}(p))  -
    \mathcal K^{(2)}{}'(\zeta_{_\mathrm{L}}(p))\Big) \Psi_t(\zeta_{_\mathrm{H}}(p)) = 0 
\end{align}
Evaluating the functional derivative explicitly yields
\begin{align}
    \mathcal K^{(2)}{}'(\zeta_{_\mathrm{H}}(p)) &=  K_2(p,p;t) \, f(p)\varphi_{_\mathrm{H}}(p) 
    \\ &\phantom{=}+ \int_{p'} \mathcal X_{\Sigma_t^p } K_2(p,p';t)f(p')\varphi(p') .\nonumber
\end{align}
When inserted into \eqref{P2:SA_condition_1}, $\mathcal P^* \mathcal P$ admits self-adjoint extensions if the following equation holds true:
\begin{align}
     \big(\varphi_{_\mathrm{H}}(p) - \varphi_{_\mathrm{L}}(p) \big)K_2(p,p;t) \Psi_t(\zeta_{_\mathrm{H}}(p) )  = 0.
\end{align}
Since $\Psi(\varphi)\equiv0$ is not a valid element of the Hilbert space $L^2(\Gamma(\mathcal{C}_t,\myd\phi)$, this indicates that either $\varphi_{_\mathrm{H}}(p) = \varphi_{_\mathrm{L}}(p)$ or $K_2(p,p;t) = 0$. 
The first derived condition violates in general \eqref{domp} unless there exists a $p_0\in\Sigma_t$ for which $\varphi_{_\mathrm{H}}(p_0) = \varphi_{_\mathrm{L}}(p_0)\equiv0$. From the second condition we have  $K_2(p,p;t) =0 $ which imply that  $\Psi(\varphi)$ cannot be an element of $\Gamma_\text{can}$ in static \cite{hatfield2018quantum} nor dynamical spacetimes.

Since \eqref{P2:SA1} covers only trivial solutions, we focus on \eqref{P2:SA2} instead. 
Using $\Psi_t(\zeta_{_\mathrm{H}}(p)) = \alpha^*(p)\Psi_t^*(\zeta_{_\mathrm{L}}(p))$, we reduce \eqref{P2:SA2} to 
\begin{align} \label{gradient_different}
    0&= \left(\mathcal K^{(2)}{}'(\zeta_{_\mathrm{H}}(p)) + 
    (\mathcal K^{(2)}{}'(\zeta_{_\mathrm{L}}(p)))^*\right) \Psi_t(\zeta_{_\mathrm{H}}(p)) 
     \\
     &= \bigg[ 2 \text{Im} \big(K_2(p,p;t)\big) f(p)\varphi_{_\mathrm{H}}(p) \nonumber 
     \\
     &\phantom{====}- \int_{p'}\mathcal X_{\Sigma_t^p } \text{Re}\big(K_2(p',p;t)\big)f(p') \varphi(p') \bigg] \Psi_t(\zeta_{_\mathrm{H}}(p))   \nonumber
\end{align} 
where we have considered $\varphi_{_\mathrm{H}}(p) = - \varphi_{_\mathrm{L}}(p)$ and limited ourselves to $\Psi_t \in\text{Dom}(\mathcal P)$ which excludes $\Psi_t\equiv0$. 

By the same consistency requirement that $\varphi_{_\text{A}}(p)$ cannot be a functional of $\varphi(p')$ for all $p'\in \Sigma_t^p$,  we integrate out all the $\varphi(p')$ dependency in \eqref{gradient_different} with $\int_{U\setminus U_p}\myd\varphi$.
To condense the necessary requirement for self-adjoint extensions we multiply both sides with $\Psi_t(\zeta_{_\mathrm{H}}) $ before we integrate over $\varphi(p')$ for all $p' \in \Sigma_t^p$. 
Given that $\Psi_t \in \text{Dom}(\mathcal P)$, one finds 
\begin{align}
    \int_{U\setminus U_p}\hspace{-1.5em}\myd\varphi  \int_{p'}\!\!\mathcal X_{\Sigma_t^p} \, \text{Re}\big(K_2(p',p;t)\big) f(p')\varphi(p')    \lvert \Psi_t(\zeta_{_\mathrm{H}}(p)) \rvert^2 = 0 \,.
\end{align}
As such we are left with the necessary condition 
\begin{align}
    \text{Im}\big(K(p,p;t)\big) \varphi_{_\mathrm{H}}(p)\, \int_{U\setminus U(p)} \myd\varphi \,  \lvert \Psi_t(\zeta_{_\mathrm{H}}(p)) \rvert^2 = 0
\end{align}
which suggests either the criterion Im$(K(p,p;t)))=0$ for all $p\in\Sigma_t$ or $\varphi_{_\mathrm{H}}(p)  =0$. If instead $\varphi_{_\mathrm{H}}(p)\neq 0 $ and $\text{Im}(K(p,p;t)) \neq 0$, we conclude $\| \Psi_t\|_{\partial U_p} = 0$. 
This can be met if either $\Psi_t$ is compact, such that $\Psi_t (\zeta_\text{A}(p)) = 0 \  \forall p\in \Sigma_t$, or $U_p$ is non-compact, such that the asymptotic fall-off conditions for $\Psi_t$ guarantee $\Psi_t (\zeta_\text{A}(p)) = 0 \  \forall p\in \Sigma_t$. 
However, we are in neither case because $\Psi_t$ is Gaussian and $U_p$ is compact. We are left with $\text{Im}(K(p,p;t)) = 0$ for all $p\in \Sigma_t$. 
Thus this is the criterion for $\mathcal P^*\mathcal P$ to admit self-adjoint extensions.

%%% Dynamics

The subsequent spectral analysis immediately trickles down to the Hamilton operator. Let $\Psi_t(\varphi) \in \text{Dom}(\mathcal P)$ a Gaussian wave-functional, its most general form is $\Psi_t(\varphi) = \mathcal{N}_t \mathcal F_t(\varphi)$  as in \eqref{guassian_exp_functional} with $\kappa = -1$. 
For $\Psi_t(\varphi)$ to be associated with physical systems, $\Psi_t(\varphi) \in \Gamma_{\rm can}$ is a solution to the functional Schr\"odinger equation
\begin{align} \label{Schrodinger_Eqt}
    \mbox{i} \partial_t \Psi_t(\varphi) =\mathcal{H}(\mathcal P, \Phi) \Psi_t(\varphi) 
\end{align}
where $\Phi$ is defined as in Section \ref{sec:ka}. 
Adopting the foliation \eqref{admmetric}, the Hamilton operator decomposes into $\mathcal{H}(\mathcal P, \Phi) =\mathcal{H}_\perp+\mathcal{H}_\myparallel$ where $\mathcal{H}_\perp=(1/2)(\mathcal{A}(\mathcal P) + \mathcal{V}(\Phi))$ and $\mathcal{H}_\myparallel= \mathcal{D}(\mathcal P,\Phi)$ where the individual terms are  \cite{Long:1996wf,Halliwell:1991ef}:
\begin{align} 
   \mathcal A(\mathcal P) &=  (f,N_\perp\mathcal T_{\varphi })
    \\ 
  \mathcal  V(\Phi) &=   (f,N_\perp\Phi (-\Delta +m^2 +\xi R) \Phi) \nonumber
    \\
  \mathcal  D(\mathcal P,\Phi) &=   ( f, N_\myparallel(\nabla \Phi) \mathcal P_{\varphi }) \nonumber
\end{align}
where $\mathcal{T}_\varphi$ is defined as in \eqref{funcLap}, the Laplace operator $\Delta = q(\nabla,\nabla)$ with metric $q$ induced on $\Sigma_t$. For $\xi$ is the coupling parameter to the Ricci scalar curvature $R$, we assume that the operator $\Delta -m^2 -\xi R$ acting on $\varphi$ yields real eigenvalues, such that the self-adjointness of $\Phi$ passes on to $\mathcal{V}(\Phi)$. 
The off-diagonal part $\mathcal D(\mathcal P,\Phi)$ is not particularly important because it can be eliminated by a (local) coordinate transformation to diagonalize the Hamilton operator. What would always remain, is the diagonal part of $\mathcal H$, that is $\mathcal A(\mathcal P) +  \mathcal V(\Phi)$.  

As a result, to conclude our analysis for $\mathcal{H}$, we are required to concern only $\mathcal{A}( \mathcal P)$, which implies to determine whether or not $\mathcal{T}_\varphi=\mathcal P^*\mathcal P$ admits self-adjoint extensions. 
To this aim we show that for a general $\mathcal K^{(2)}$ satisfying \eqref{Schrodinger_Eqt}, $\mathcal{H}$ features complex functionals of $\varphi$ if the space-time is dynamical, such that $\mathcal{H}$ cannot admit any self-adjoint extension, although they exist for $\mathcal{P}$. Let us sketch the solution space $\Gamma_{\rm can}$ determined by \eqref{Schrodinger_Eqt} for a free scalar field theory on a dynamical, globally hyperbolic space-time. For the Gaussian state  $\Psi_t(\varphi)=\mathcal{N}_t\mathcal{F}_t(\varphi)$ we find the following two equations to be satisfied by $\mathcal{K}^{(2)} $ 
\begin{align}
    -\mbox{i}\partial_t \mathcal K^{(2)} &=\Big(f,N_\perp\big( (1/2) \mathcal P_{\varphi}\, \mathcal K^{(2)} (\varphi) \big)^2 + N_\perp V(\varphi)\nonumber
    \\ &\phantom{====} -
      N_\myparallel(\nabla \Phi) \mathcal P_{\varphi} \,\mathcal K^{(2)}(\varphi)\Big)\label{oscpart}
    \\
    \mbox{i}\partial_t \ln (\mathcal{N}_t) &=- \frac 1 2 \big(f, N_\perp
    \mathcal T_\varphi \mathcal{K}^{(2)}(\varphi)\big),\label{normpart}
\end{align}
where $V(\varphi)=\varphi(-\Delta+m^2+\xi R)\varphi$. All $\Psi_t$ that solve the above equations span $\Gamma_{\rm can}$. It is obvious that any space-time dependence comes from the metric $g$ which is implicitly contained in $\mathcal{V}(\varphi)$, $N_\perp$, $N_\myparallel$, etc. Now we show that these elements are incompatible with the self-adjoint domain of $\mathcal H$, Dom$(\mathcal H)$, by contradiction.
Recall, the self-adjoint domain requires Im$(K_2(p,p;t))=0$,
consider now the generic solution to \eqref{normpart} in local description \cite{Long:1996wf}
\begin{align}\label{nint}
   \frac{\lvert \mathcal{N}_t \rvert^2}{\lvert \mathcal{N}_{t_0}\rvert^2}= \exp\bigg[ \mbox{i} \int_{p} \int_{t'}   N_\perp \, \text{Im} \big(K_2(p,p;t')\big) \bigg],
\end{align}
where $\mathcal{N}_{t_0}$ denotes the integration  constant of the $t'$-integral. 
Suppose $\text{Im}(K_2(p,p;t')) = 0$ for all $t'\in[t_0,t]$, then  $\lvert \mathcal{N}_t \rvert^2 = \lvert \mathcal{N}_{t_0} \rvert^2$.
Since $\text{Im}(K_2(p,p;t)) = 0$, $\mathcal{H}$ admits a self-adjoint extension which ensures norm conservation $\partial_t\|\Psi_t\| = 0$. By direct computation, we find  
\begin{align}\label{contra}
   \frac{\partial_t \|\Psi_t\|}{\lvert \mathcal{N}_{t_0} \rvert^{2}} = - \int_U \myd\varphi \ \partial_t \text{Re}\big(\mathcal K^{(2)}\big) \exp\left(-\text{Re}\big(\mathcal K^{(2)}\big)\right)= 0.
\end{align} 
This identity is true for all $p\in \Sigma_t$, if and only if $\partial_t \text{Re}(\mathcal K^{(2)}) = 0$ for $ \text{Re}(\mathcal K^{(2)}) \neq 0$, or otherwise \eqref{contra} is false. 
The former requirement $\partial_t \text{Re} (\mathcal K^{(2)}) = 0$ is possible only if the kernel is time independent which is tantamount to saying that the space-time is static.
In this case, \eqref{Schrodinger_Eqt} reduces to the time independent Schr\"odinger equation $\mathcal H(\mathcal P,\Phi) \Psi_t(\varphi) = E \,\Psi_t(\varphi)$ for $E\in \mathbb R$ \cite{hatfield2018quantum}. 
If, however, the space-time is dynamical, then $\partial_t\text{Re}(\mathcal K^{(2)}) \neq0 $ implies that by consistency  it is impossible to meet $\text{Im}(K_2(p,p;t)) =0$   such that the  Hamilton operator has a self-adjoint extension.
On the contrary, when a space-time is static, for example, Minkowski space-time, then it is clearly possible for $\mathcal{H}$ to admit infinitely many self-adjoint extensions whilst $\Psi_t(\varphi)\in\Gamma_{\rm can}$.

\section{Contractive Evolutions} 
\label{sec:ce}
Until now, we only considered the spectrum of $\mathcal{H}$ which eventually generates the time-evolution operator. As shown in the previous section, $\mathcal{H}$ is not necessarily essentially self-adjoint. Here, we delve deeper into the implications and consequences for the evolution group. This allows us to morph Stone's pair in quantum mechanics (essentially self-adjoint Hamilton operator $\Leftrightarrow$ unitary evolution group) into a generalized, functional version.

For reasons that will soon become clear, 
we consider a one-parameter family of evolution operators
$\{\mathcal{E}(t,t_0)\}_{t\in I}$ on 
$L^2(\Gamma(\mathcal{C}_{t_0}),\myd\phi)$
(with $\mathcal{C}_{t_0}$ denoting the initial configuration space  
at time $t_0\in I$)
satisfying 
$\mathcal{E}(t_0,t_0)=
\mathrm{id}_{\Gamma^*(\mathcal{C}_{t_0})}$,
the map $t\mapsto \mathcal{E}(t,t_0) \Psi_\mathrm{in}$
is continuous for each initial wave functional $\Psi_\mathrm{in}$
in $L^2(\Gamma(\mathcal{C}_{t_0}),\myd\phi)$ and $t\in I$. Moreover, 
we consider only a special family of evolution operators, 
called contractive \cite{ree75}, which satisfy the additional condition
\begin{eqnarray}
   && \mathrm{inf}\Big\{
    C\ge 0:
    \mathbb{E}^{1/2}\left(\mathrm{id}_{\Gamma^*(\mathcal{C}_t)} \, ; \,
    \mathcal{E}(t,t_0)\Psi_\mathrm{in} \right)
    \le
    C \; 
    \nonumber \\
    && \hspace{2.4cm}\mathrm{for \; all} \; 
    \Psi_\mathrm{in} \; \mathrm{in} \; 
    L^2(\Gamma(\mathcal{C}_{t_0}),\myd\phi)
    \Big\} 
    \le 1
    \nonumber
\end{eqnarray}
Contractive evolution families arise naturally 
in effective field theories
and provide a necessary generalization of the relationship between 
unitary evolution groups and self-adjoint generators of 
time translations. 

As in the case of contracting evolution semigroups
(including unitary representations of evolution groups), 
we obtain the generator of $\mathcal{E}(t+\delta t, t)$
for $[t,t+\delta t]\subset I$ as follows:
Introduce\footnote{Of course, $T_t$ will be the Hamilton operator $\mathcal H$ but since this is a more general analysis, we decided to rename the generator for time translations.}
$
    T_{t}(\delta t)
    :=
    (\delta t)^{-1}
    (
        \mathrm{id}_{\Gamma^*(\mathcal{C}_t)} 
        - \mathcal{E}(t+\delta t,t)
    )
$
and consider the domain 
$
    \mathrm{Dom}(T_t)
    :=
    \{\Psi_t\in L^2(\Gamma(\mathcal{C}_t),\myd\phi): 
    \lim_{\delta t\rightarrow 0^+} T_t(\delta t) \Psi_t 
    \; \mathrm{exists}\}
$
for the infinitesimal generator $T_t$, defined by 
$T_t \Psi_t := \lim_{\delta t\rightarrow 0^+}T_t(\delta t) \Psi_t$
for all $\Psi_t$ in $\mathrm{Dom}(T_t)$.
We will also say that $T_t$ generates $\mathcal{E}(t+\delta t,t)$
and write the time-ordered exponential
\begin{eqnarray}
    \mathcal{E}(t,t_0)
    &=&
    \mathcal{T} \exp{\left(-
    \int_{[t_0,t]} \mathrm{d}\tau \;  T_\tau(\Phi,\mathcal{P})
    \right)}
    \; .
\end{eqnarray}
Note that $\mathrm{Dom}(T_t)$ is dense in $L^2(\Gamma(\mathcal{C}_t),\myd\phi)$:
consider any $\Psi_t$ in $L^2(\Gamma(\mathcal{C}_t),\myd\phi)$, 
and set 
\begin{eqnarray}
    \Psi^{(s)}(\phi) 
    := 
    \int_{[0,s]} \mathrm{d}\tau \; 
    \mathcal{E}(t+\tau,t) \Psi_t(\phi)
\end{eqnarray}
for $[t,t+s]\subset I$. Here, the idea is that 
the sequence $\{s^{-1} \Psi^{(s)}\}_{s\in I}$
converges to $\Psi_t$ as $s$ approaches zero. 
If $\Psi^{(s)}$ is in $\mathrm{Dom}(T_t)$ for any $s>0$
and for each $\Psi_t\in$ 
$L^2(\Gamma(\mathcal{C}_t),\myd\phi)$, then any $\Psi_t$
is the limit of a sequence in $\mathrm{Dom}(T_t)$, and, thus, $\mathrm{Dom}(T_t)$ 
would be dense in $L^2(\Gamma(\mathcal{C}_t),\myd\phi)$. 
Now, for any $r>0$ such that $(t+s+r)\in I$, consider 
$\mathcal{E}(t+r,t) \, \Psi^{(s)}$. 
We can rewrite $\mathcal{E}(t+r,t) \, \Psi_{t+\tau}$
as $\Psi_{t+\tau+r}$, because the evolution operator 
reduces to an ordinary exponential describing time-translation 
when it is applied to a solution of the evolution equation
$\partial_t \Psi_t = - T_t \Psi_t$.
Hence, 
\begin{eqnarray}
    T_t (r) \Psi^{(s)}
    &=&
    r^{-1} 
    \int_{[0,s]}
    \mathrm{d}\tau 
    \left(
        \Psi_{t+\tau} - \Psi_{t+\tau+r}
    \right)
    \nonumber \\
    &=&
    r^{-1}
    \left(
        \int_{[0,s]} \mathrm{d}\tau \, \Psi_{t+\tau} 
        - \int_{[r,r+s]} \mathrm{d}\tau \,  \Psi_{t+\tau} 
    \right)
    \nonumber \\
    &\longrightarrow&
    \left(\mathrm{id}_{\Gamma^*(\mathcal{C}_t)} - \mathcal{E}(s+t,t) \right)\Psi_{t}
    \hspace{0.2cm} \mathrm{for} \; r\rightarrow 0 \; . 
\end{eqnarray}
So for each $\Psi_t$ in $L^2(\Gamma(\mathcal{C}_t),\myd\phi)$
and for each admissible $s>0$, the sequence $\{s^{-1}\Psi^{(s)}\}_{s\in I}$ 
lies indeed in $\mathrm{Dom}(T_t)$, and, therefore, $\mathrm{Dom}(T_t)$ is dense in 
$L^2(\Gamma(\mathcal{C}_t),\myd\phi)$.
It is worth stressing that the evolution equation is required in this 
derivation since, in general, the dynamical content of the theory 
is given by a one-parameter family of time-ordered exponentials
which does not even constitute a semi-group. 
However, if $\Psi_t$ is in the domain of $T_t$, then 
$T_{s+t} \, \mathcal{E}(s+t,t) \,\Psi_t$
$=$
$\mathcal{E}(s+t,t) \, T_t \, \Psi_t$, so 
$\mathcal{E}(s+t,t):$
$\mathrm{Dom}(T_t) \rightarrow \mathrm{Dom}(T_{s+t})$, and 
$\partial_s \Psi_{s+t}$
$=$
$- T_{s+t} \Psi_{s+t}$ 
$=$
$-(T_t \Psi_t)_{s+t}$.

We can use this to show that $T_t$ is also closed:
Let $\{\Psi_t^n\}_{n\in\mathbb{N}}$ be a sequence of 
wave functionals in $\mathrm{Dom}(T_t)$ converging to $\Psi_t$
in $L^2(\Gamma(\mathcal{C}_t),\myd\phi)$, and 
$T_t \Psi_t^n \rightarrow \Xi_t$. Then 
$T_t(r) \Psi_t$
$=$
$\lim_{n\rightarrow \infty}$
$r^{-1}(\Psi_t^n - \mathcal{E}(r+t,t)\Psi_t^n)$.
Integrating the evolution equation gives
$\mathcal{E}(r+t,t)\Psi_t^n$
$=$
$\Psi_t - (T_t\Psi_t^n)^{(r)}$, where 
$T_{s+t} \, \mathcal{E}(s+t,t) \,\Psi_t^n$
$=$
$\mathcal{E}(s+t,t) \, T_t \, \Psi_t^n$
has been used (since the sequence is 
by assumption in the domain of $T_t$). 
Thus, 
$T_t(r) \Psi_t$
$=$
$\lim_{n\rightarrow\infty}$
$r^{-1}$
$(T_t\Psi_t^n)^{(r)}$. 
The limit is $r^{-1}\Xi^{(r)}$. 
For $r\rightarrow 0$, we find 
$T_t(r) \Psi_t \rightarrow \Xi_t$, so $\Psi_t \in \mathrm{Dom}(T_t)$
and $T_t\Psi_t = \Xi_t$. Since $T_t$ is closed, this allows to perform a generalized spectral analysis which is essential to relax the description of Stone's pair. Now we turn our attention to the necessary property that is attributed to the generator.

\subsection{Accretive generators}
\label{sec:acc}
We proceed with considering 
contracting evolution operators and remind the reader 
that they are a (necessary) generalization of unitary 
evolution operators. While the contraction property 
replaces unitarity, it remains to investigate 
what supplants self-adjointness as the related
property of the infinitesimal generators. 

%%% GENERAL I LATER

Let $E_s\in\mathbb{C}$ be in the spectrum of 
the infinitesimal generator $T_s$, $s\in I=(0,\infty)$ (for simplicity). 
Introduce 
$\gamma_s := -\partial_s$ and 
$G_s(t):= \exp{(-t \gamma_s)}$. 
Consider the formal Laplace transform\footnote{Considering a one-sided exponential decay multiplied by a Heaviside distribution $\Theta(x)$ which resembles a contracting semi-group evolution. The corresponding Laplace transform yields $\mathcal{L}[e^{-ax}\Theta(x)](s)=(a+s)^{-1}$.}  
\begin{equation}
    \mathcal{L}[G_s(t)](E_s)=-(E_s \, \mathrm{id}_{\mathrm{Dom}(T_s)}+ \gamma_s)^{-1}.
\end{equation}
Assume that $\mathrm{Re}(E_s)>0$. Then,  
if $\Psi_s=\mathcal{E}(s,0)\Psi_0$, and since 
$\|G_s(t)\|\le 1$ by assumption, 
$\mathcal{S}\Psi_s$
$=$
$\mathcal{L}[G_s(t)\Psi_s](E_s)$
is a bounded linear operator of norm less or equal 
to $(\mathrm{Re}(E_s))^{-1}$.
For positive $\delta t$, 
\begin{eqnarray}
    &&\gamma_s(\delta t) \, \mathcal{S}\Psi_s
    =
    \nonumber \\
    &&=
    \frac{1}{\delta t}
    \left(\,
        \mathcal{L}[\left(G_s(t) - G_s(t+\delta t)\right) \Psi_s](E_s)\,
    \right) \\
    &&=
    \left(\frac{1-\mathrm{e}^{\delta t E_s}}{\delta t}\right)
    \mathcal{L}[\Psi_t](E_s)+
    \frac{\mathrm{e}^{\delta t E_s}}{\delta t}
    \int_{[0,\delta t]} 
    \mathrm{e}^{ t E_s} \, \Psi_t
    \; , \nonumber 
\end{eqnarray}
where 
the second equality holds after shifting $t\rightarrow t-\delta t$
in the second Laplace transform, and $\Psi_t := G_s(t)\Psi_s$.
In the limit $\delta t\rightarrow 0$, we find 
$\gamma_s(\delta t) \, \mathcal{S}\Psi_s$ 
$\rightarrow$
$-E_s \mathcal{S} \Psi_s+\Psi_s$.  
Hence, $\mathcal{S}\Psi_s\in \mathrm{Dom}(\gamma_s)$ and 
$\gamma_s \mathcal{S} \Psi_s = \Psi_s - E_s \mathcal{S}\Psi_s$.
This implies
$
(E_s \mathrm{id}_{\mathrm{Dom}(T_s)}+\gamma_s) \mathcal{S} \Psi_s = \Psi_s
$.
Furthermore, for $\Psi_s\in \mathrm{Dom}(T_s)$, we have 
$ 
    \gamma_s \mathcal{S} \Psi_s
    =
    \mathcal{L}(\gamma_s \Psi_t)(E_s)
    =
    \mathcal{L}(G_s(t)\gamma_s \Psi_s)
$, so $[\gamma_s,\mathcal{S}]=0$ on $\mathrm{Dom}(T_s)$
since $\mathrm{e}^{-\delta t E_s} \Psi_t$ and
$\gamma_s \mathrm{e}^{-\delta t E_s} \Psi_t$
are integrable by the condition on the spectrum of $\gamma_s$, 
and the fact that $\gamma_s$ is closed. As a consequence, 
for $\Psi_s\in \mathrm{Dom}(T_s)$, the following holds:
$
\mathcal{S}(E_s \mathrm{id}_{\mathrm{Dom}(T_s)} + \gamma_s) \Psi_s 
=
(E_s \mathrm{id}_{\mathrm{Dom}(T_s)}+\gamma_s) \mathcal{S} \Psi_s = \Psi_s
$, which implies that 
$
\mathcal{S} = (E_s \mathrm{id}_{\mathrm{Dom}(T_s)}+\gamma_s)^{-1}
$
holds in the strong sense. 

Apart from the restrictive adaption to include time-ordered 
exponentials as evolution operators, the above reasoning 
follows the proof of the necessity part of the Hille-Yosida 
theorem \cite{ree75}. 
The spectral properties of $T_s$ are also sufficient 
to guarantee a contracting family of evolution operators:
let $E_s$ be real positive and introduce 
$\vartheta(E_s)$
$:=$
$E_s (\,\mathrm{id}_{\mathrm{Dom}(T_s)}$ 
$-$ 
$E_s (E_s \mathrm{id}_{\mathrm{Dom}(T_s)} + \gamma_s)^{-1}\,)$
on $\mathrm{Dom}(T_s)$.
The derivation proceeds in three steps: 
first, we will prove that 
$\vartheta(E_s)\Psi_s \rightarrow \gamma_s \Psi_s$ 
as $E_s\rightarrow\infty$
for any wave functional in $\mathrm{Dom}(T_s)$. Then we will show 
that the semigroups $\exp{(-t \vartheta(E_s)})$ 
are contracting and construct $G_s(t)$ as the strong limit
of these semigroups. 

For $\Psi_s=\mathcal{E}(s,0) \Psi_0$ in $\mathrm{Dom}(T_s)$, 
we have 
$\vartheta(E_s) \Psi_s$
$=$
$E_s(E_s\, \mathrm{id}_{\mathrm{Dom}(T_s)}+\gamma_s)^{-1}\gamma_s \Psi_s$.
Moreover, from the necessity part discussed above, 
$\|(E_s\, \mathrm{id}_{\mathrm{Dom}(T_s)}+\gamma_s)^{-1}\|\le E_s^{-1}$
for all real positive $E_s$, so 
\begin{eqnarray}
    - E_s^{-1} \vartheta(E_s) \Psi_s
    &=&
    E_s(E_s\, \mathrm{id}_{\mathrm{Dom}(T_s)}+\gamma_s)^{-1} \Psi_s
    - \Psi_s
    \nonumber \\
    &=&
    - (E_s\, \mathrm{id}_{\mathrm{Dom}(T_s)}+\gamma_s)^{-1}\gamma_s \Psi_s
    \stackrel{E_s\rightarrow\infty}{\longrightarrow} 0 \; ,
    \nonumber
\end{eqnarray}
since 
$
    \|E_s^{-1} \vartheta(E_s) \Psi_s \|
    =
    \|(E_s\, \mathrm{id}_{\mathrm{Dom}(T_s)}+\gamma_s)^{-1}\gamma_s \Psi_s\|
    \le
    E_s^{-1} \|\gamma_s \Psi_s\|
$
by the above bound. 
It follows that the family
$ 
    \{(E_s\, \mathrm{id}_{\mathrm{Dom}(T_s)}+\gamma_s)^{-1}: E_s>0\}
$ 
is uniformly bounded in norm, so 
$(E_s\, \mathrm{id}_{\mathrm{Dom}(T_s)}+\gamma_s)^{-1}\Psi_s$
$\rightarrow$
$\Psi_s$ for all $\Psi_s$ in
$L^2(\Gamma(\mathcal{C}_s),\myd\phi)$, given that $\mathrm{Dom}(T_s)$
is dense in this space. 
Hence,
$\vartheta(E_s) \Psi_s$
$=$
$E_s(E_s\, \mathrm{id}_{\mathrm{Dom}(T_s)}+\gamma_s)^{-1}\gamma_s\Psi_s$
$\rightarrow$ $\gamma_s \Psi_s$ as $E_s\rightarrow \infty$. 

The semigroups with infinitesimal generators $\vartheta(E_s)$ 
are defined by power series. Since
\begin{eqnarray}
    &&\left\|
        \mathrm{exp}\left(-t\vartheta(E_s)\right)
    \right\|
    =
    \nonumber \\
    &&=
    \left\|
        \mathrm{exp}\left(-t E_s\right)
        \mathrm{exp}\left(+t E_s^2(E_s \, \mathrm{id}_{\mathrm{Dom}(T_s)} 
        + \gamma_s)^{-1}\right) 
    \right\| 
    \nonumber \\
    &&\le
    \mathrm{exp}\left(-t E_s\right)
    \sum_{n\in\mathbb{N}}
    \frac{(t E_s^2)^n}{n!}
    \left\|(E_s\, \mathrm{id}_{\mathrm{Dom}(T_s)}+\gamma_s)^{-1}\right\|^n
    \nonumber \\
    &&\le
    1 \; , \nonumber
\end{eqnarray}
where the first equality is due to the definition of $\vartheta(E_s)$, 
the first equality follows from the triangle inequality 
and the last from the bound on $\|(E_s\, \mathrm{id}_{\mathrm{Dom}(T_s)}+\gamma_s)^{-1}\|$, 
because they are contracting semigroups. 
Let $E_s, {E_s}^\prime, t$ be real positive, and $\Psi_s\in \mathrm{Dom}(T_s)$. Then 
\begin{eqnarray}
    &&\left(
        \mathrm{exp}\left(-t \vartheta(E_s)\right)
        -
        \mathrm{exp}\left(-t \vartheta({E_s}^\prime)\right)
    \right)\Psi_s =
    \nonumber \\
    &&=
    \int_{[0,t]} \mathrm{d} x \;
    \frac{\mathrm{d}}{\mathrm{d} x}
    \left(
        \mathrm{exp}\left(-x \vartheta(E_s)\right)
        \mathrm{exp}\left(-(t-x) \vartheta({E_s}^\prime)\right) \Psi_s
    \right)
    \nonumber \; .
\end{eqnarray}
Using that $\{\vartheta(E_s)\}_{E_s>0}$ is a commuting 
family of infinitesimal generators, 
\begin{eqnarray}
    &&\left\|
     \left(  \mathrm{exp}\left(-t \vartheta(E_s)\right)
        -
        \mathrm{exp}\left(-t \vartheta({E_s}^\prime)\right)
    \right)\Psi_s  
    \right\|
    \nonumber \\ 
    &&\le
    \int_{[0,t]} \mathrm{d} x \;
    \left\|
        \mathrm{exp}\left(-x \vartheta(E_s)\right)
        \mathrm{exp}\left(-(t-x) \vartheta({E_s}^\prime)\right)
    \right\|\times
    \nonumber \\
    &&\hspace{3.8cm}
  \times  \left\|
        \vartheta\left({E_s}^\prime\right) \Psi_s
        -
        \vartheta\left(E_s\right) \Psi_s
    \right\|
    \nonumber \\ 
    &&\le
    t 
    \left\|
       \vartheta\left({E_s}^\prime\right) \Psi_s
        -
        \vartheta\left(E_s\right) \Psi_s 
    \right\| 
    \; . \nonumber
\end{eqnarray}
The last inequality follows from the contraction property
of the semigroups generated by $\vartheta(E_s)$. 
Since we have shown above that $\vartheta(E_s)\Psi_s$
converges to $\gamma_s \Psi_s$ as $E_s\rightarrow \infty$, 
$(\mathrm{exp}(-t \vartheta(E_s))\Psi_s)_{E_s>0}$
is a Cauchy sequence in this limit for any real positive $t$
and $\Psi_s$ in $\mathrm{Dom}(T_s)$.
Let us set
$
G_s(t)\Psi_s
:= 
\lim_{E_s\rightarrow\infty}\mathrm{exp}(-t \vartheta(E_s)) \Psi_s
$, since the properties of contraction semigroups are preserved 
under strong limits, $G_s(t)$ constitutes a semigroup of contracting 
evolution operators. The above inequality shows that $G_s(t)$
is a strongly continuous contraction semigroup. 

Let $\tilde{\gamma}_s$ be the infinitesimal generator of $G_s(t)$. 
For all $t>0$ and $\Psi_s\in\mathrm{Dom}(T_s)$, 
\begin{eqnarray}
    &&\left(
        \mathrm{id}_{\mathrm{Dom}(T_s)} - \mathrm{exp}(-t \vartheta(E_s))
    \right) \Psi_s
    =
    \nonumber \\
    &&=
    \int_{[0,t]} \mathrm{d}x \; 
    \mathrm{exp}(-x \vartheta(E_s))
    \vartheta(E_s) \Psi_s
    \; , \nonumber
\end{eqnarray}
and therefore 
\begin{eqnarray}
    \left(\mathrm{id}_{\mathrm{Dom}(T_s)}- G_s(t)\right) \Psi_s
    &=&
    \int_{[0,t]} \mathrm{d}x \; 
    G_s(x) \gamma_s \Psi_s  \nonumber
\end{eqnarray}
since $\vartheta(E_s)\Psi_s$ converges to $\gamma_s \Psi_s$ 
in the limit $E_s\rightarrow\infty$.
Now, in the limit $t\rightarrow 0^+$, the left-hand side 
converges to $\tilde{\gamma_s}\Psi_s$ and the right-hand side 
to $\gamma_s\Psi_s$. Thus 
$\tilde{\gamma}_s \Psi_s$ converges to $\gamma_s \Psi_s$. 
Therefore $\mathrm{Dom}(\tilde{\gamma}_s)\supset \mathrm{Dom}(\gamma_s)$
and $\tilde{\gamma}_s$ restricted to the domain $\mathrm{Dom}(\gamma_s)$
agrees with $\gamma_s$. It remains to show that both domains 
coincide. For real positive $E_s$, 
the inverse of $(E_s \mathrm{id}_{\mathrm{Dom}(T_s)} + \tilde{\gamma}_s)$
exists by the necessity part of the statement shown above, 
and the inverse of $(E_s \mathrm{id}_{\mathrm{Dom}(T_s)} + \gamma_s)$
exists by hypothesis. Hence, 
$(E_s \mathrm{id}_{\mathrm{Dom}(T_s)} + \tilde{\gamma}_s)\mathrm{Dom}(\tilde{\gamma}_s)$
$=$
$L^2(\Gamma(\mathcal{C}_s), \myd\phi)$, and, as well, 
$(E_s \mathrm{id}_{\mathrm{Dom}(T_s)} + \gamma_s)\mathrm{Dom}(\gamma_s)$
$=$
$L^2(\Gamma(\mathcal{C}_s), \myd\phi)$, so indeed
$\mathrm{Dom}(\tilde{\gamma}_s)=\mathrm{Dom}(\gamma_s)$. 

The above derivation requires to construct the resolvent of 
the infinitesimal generator of the evolution operator 
in order to verify the spectral properties that are necessary 
and sufficient to generate a one-parameter family of contracting 
evolution operators. If $\Psi_0$ is in 
the domain of $T_\tau$ for $\tau\in I:=[0,t]$, then 
the contraction property implies within Dom$(T_\tau)$
\begin{equation}    
-\partial_t\|\Psi_t\|
   =
    \mathbb{E}(-T_t^*; \Psi_t) + \mathbb{E}(-T_t; \Psi_t)
\end{equation}
Therefore, the contraction property requires 
that Re$(\mathbb{E}(-T_t; \Psi_t))$ is 
semi-positive definite. Such a densely defined infinitesimal 
generator is called accretive. 

By identifying the infinitesimal generator with the Hamilton operator $\mathcal{H}$, we can link this with the previous discussion. Albeit being complex-valued, if the spectrum admits a suitable sign in the imaginary (or real part), the evolution can be described by evolution operators that keep the probabilistic feature of the theory intact \cite{Hofmann:2015xga,Eglseer:2017kcs}. In the spirit of Stone's theorem we see that 
one-parameter families of evolution operators 
admit in general contracting representations 
generated by accretive operators. 
This includes the celebrated situations that 
allow for a unitary evolution generated by 
self-adjoint operators which  
are formal observables of the theory \cite{Ashtekar:1975zn}. 

\subsection{Initial data}
\label{sec:id}
In sections \ref{sec:da}, \ref{sec:acc}, statements concerning 
the dynamical content of the theory have been derived 
assuming time-evolved data as input. Consequently these 
statements do not extend to wave functionals on Cauchy 
hypersurfaces. 
The goal of this section is to close this loophole.

For simplicity, let $I$ be a half-open, left-closed interval
of the extended real numbers with $t_0:=\partial I=0$
denoting its minimum element, and $J:=[\partial I, \delta t]\subset I$ 
with $\delta t$ being an arbitrary small positive number
such that $\Psi_{\tau}:=\mathcal{E}(\tau,0) \Psi_0$
is continuous on $J$ and smooth on the interior of $J$. 
Consider the integrated evolution equation
\begin{eqnarray}
    \Psi_{\delta t}
    &=&
    \Psi_0 - \int_J \mathrm{d}s \; T_s \Psi_s
    \; .
\end{eqnarray}
The mean value theorem for definite integrals guarantees 
that there is a real number $m$ 
in the interior of $J$ such that 
$
    \Psi_{\delta t}
    =
    \Psi_0 - \delta t \, T_m \Psi_m 
$.
We can write $m=\alpha \delta t$ with $\alpha \in (0,1)$. 
Since $m$ is a positive fraction of $\delta t$, it follows 
that $\Psi_{\delta t} = \Psi_0 - \delta t\, T_{0^+} \Psi_0 + O(\alpha)$, 
where $0^+$ is in the interior of $J$ arbitrary close 
to the minimum element of $I$, and $O$ is the Bachmann-Landau notation 
to denote the order of approximation. 
This can be seen as follows: 
by iteration of the integrated evolution equation, 
$
    \Psi_{\delta t}
    =
    \Psi_0 -\delta t \, T_{m_1} \Psi_0 + m_1 \delta t \, 
    T_{m_1} T_{m_2} \Psi_0 
    + O(\alpha^2)
$, where the given order of approximation refers to the product 
$\alpha_1 \alpha_2$ with each factor parameterizing 
the numbers $m_1>m_2$ guaranteed by the mean value theorem.
In the limit $\alpha_1 \rightarrow 0^+$, we find 
$\Psi_{\delta t} = \Psi_0 - \delta t\, T_{0^+} \Psi_0$. 
Since $\{\alpha_a\}_{a\in\mathbb{N}}$ is a strictly decreasing  
sequence (again by the mean value theorem), the 
definition of the time-ordered evolution operator 
shows that the above result holds to arbitrary precision. 
The infinitesimal 
evolution operator $(\mathrm{id}_{\mathrm{Dom}(T_{0^+})}- \delta t \, T_{0^+})$
is (at linear order in small quantities) a member of a one-parameter
family of strongly continuous semigroups: the existence of an identity 
is obvious, the composition law holds at linear level and the 
continuity property by hypothesis. 

\section{Cosmological Space-times}\label{sec:cst}

In this section, we translate our previous analysis to applications in cosmological space-times through physically relevant examples that involves an effective configuration space.
We consider two setups: a radiation-dominated universe and the Poincar\'e patch of de Sitter space-time which, both, are widely used in cosmology. As a proof of concept, we analyze how a minimally coupled scalar quantum field is being amplified to the extent that the semiclassical approximation breaks down.

\subsection{Setup} 
Consider a conformally flat Friedmann-Lema\^itre-Robertson-Walker space-time with metric 
\begin{equation}\label{FLRW:st}
    g=a^2(\eta)\left(-\mathrm{d}\eta\otimes\mathrm{d}\eta+\mathrm{d}\mathbb{E}_3\right)
\end{equation}
where d$\mathbb{E}_3$ denotes the line element of the three-dimensional Euclidean space-time and $a(\eta)$ the scale factor.
Let $\Psi_\eta(\varphi) \in \Gamma_{\rm can}$ be a Gaussian wave-functional in the solution space of \eqref{Schrodinger_Eqt} with metric \eqref{FLRW:st}. The corresponding probability density functional is given by
\begin{align}
   \rho_\eta(\varphi) = \lvert \Psi_\eta(\varphi) \rvert^2 = \lvert \mathcal{N}_\eta \rvert^2 \, \exp\Big( - \text{Re}\big(\mathcal K^{(2)}(\varphi)\big) \Big) 
\end{align}
where $\mathcal K^{(2)}(\varphi) $ fully determines the Gaussian statistics of $\rho_\eta(\varphi)$ \cite{Hofmann:2015xga, Hofmann:2019dqu, Eboli:1988qi, Long:1996wf}. Since, \eqref{FLRW:st} admits a Euclidean line element, we can perform a spatial decomposition of $\mathcal K^{(2)}$ into Fourier modes to extract the variance 
$\sigma(k;\eta)$ -- for convenience in the ultralocal description
\begin{align}
    & \mathcal K^{(2)}(\varphi)= \iint_{k,k'}\varphi(\vec k) \sigma^{-2}(k;\eta)  \delta^3(\vec k - \vec k') \varphi(\vec k') 
\end{align}
where we have identified the variance to be $\sigma^2(k;\eta) \equiv (  \mathrm{det}(q) \text{Re}(\hat K_2(
k;\eta)) )^{-1}$. 
The kernel function $\hat K_2(k;\eta)$  satisfies \eqref{Schrodinger_Eqt} and yields the solution \cite{Long:1996wf}
\begin{align}
    \hat K_2(k;\eta) = - \frac{\mbox{i}}{N_\perp\sqrt{\mathrm{det}(q)} } \partial_\eta \ln \left(\frac{u_k^*(\eta)}{u_k^*(\eta_0)}\right) ,
\end{align}
where $u_k(\eta)$ is the mode function that solves the Klein-Gordon equation for \eqref{FLRW:st}.
The modes $u_k(\eta)$ are furthermore normalized with respect to the Wronski determinant 
$
u_k(\eta) \partial_\eta u_k^*(\eta) - u_k^*(\eta) \partial_\eta u_k(\eta) = \mbox{i}N_\perp/\sqrt{\mathrm{det}(q)} 
$
\cite{Parker:2009uva}, such that, the variance can be recast into the familiar form 
$\sigma^2(k;\eta)= \lvert{u_k(\eta)}\rvert^2$ \cite{Eboli:1988qi,Parker:2009uva, Long:1996wf,Mukhanov:2007zz}. 

In particular,  $\rho_t(\varphi)$ is subjected to the following mean value and variance:
\begin{align} 
    \label{gauss:mean}
    &\mathbb E  (\mathcal X_{U_0}\cdot \Phi(k); \Psi_\eta ) = 0,  
    \\ 
    &\mathbb E  (\mathcal X_{U_0} \cdot \Phi(p)\Phi(p'); \Psi_\eta )   \approx \text{Re}\left({K}_2^{-1}(p,p';\eta)\right)\label{gauss:std} 
    \\
    &= \int^{\infty}_0 P(k,\eta) \, \frac{\sin(kr)}{kr} \, \frac{\mathrm dk}{k}\,. \nonumber
\end{align}
where $r = \lvert p-p'\rvert $ and  we have used the correlation  $\mathbb E (\mathcal X_{U_0} \cdot \Phi(k) \Phi(k'); \Psi_\eta ) =  \sigma^2(k;t) {\delta^3(k+k')}$  \cite{guth1985quantum,Eboli:1988qi,pi1990quantum} to construct the well known  power spectrum $P(k,\eta) = ( 4\pi^2)^{-1}  k^3\sigma^2(k,t)$  \cite{Mukhanov:2007zz, Parker:2009uva}.  

The mathematical consistency behind the construction of the effective configuration space concerns the study of the operator $\Phi(f) $ with respect to the boundaries set by the semiclassical approximation.
Let $\ell_p$ be a short distance cutoff, and  $m_p^{-1}=  \ell_p $ be the smallness parameters such that   $\varphi= \varphi_0 + \delta \varphi/m_p$ is assumed to be a perturbative expansion, with $\varphi=\delta\varphi/m_p$ when $\varphi_0=0$.
To quantify its validity, we consider \eqref{cond_sd} from which follows that, at least, $  \lvert \Phi(f) \rvert \in \mathcal A_0$. 
For this purpose, we choose the dimensionless ratio $\mathcal{R}(|\Phi(f) |)$ in \eqref{cond_sd} to qualify the validity of the perturbative expansion by  
\begin{equation}
    \mathcal R(\lvert \Phi(f) \rvert) := \frac{|\Phi(f)|}{m_p v_f}
\end{equation} 
where $v_{f}:=\int_{x\in\Sigma_t}f(x)$ is the coordinate volume given by supp$(f)$\footnote{This volume plays the role of spatial averaging in the above equation.}.
As a result, the filtering semi-norm is given by  
\begin{equation} \label{Filtration_for_EFT_Cosmology}
     N_{\phi_0,\Psi_\eta} (\lvert \Phi(f) \rvert) 
     := \frac{1}{m_pv_f}  \mathbb{E}(\mathcal{X}_{U_0}\cdot|\Phi(f)|;\Psi_\eta) 
\end{equation}
We considered $U_0$ as the neighborhood centered around $\phi_0 = 0$. Since $\Psi_t \in \text{Dom}(\mathcal P)$, $\partial U_0$ is symmetric in the sense that $\partial U_0(p) =\{\phi_\Lambda(p), -\phi_\Lambda(p)\}$ for all $p\in\Sigma_\eta$.
We also demand that the boundary is uniform for all $p$, i.e.  $\phi_\Lambda(p) = \phi_\Lambda(p')$ for all $p\neq p'$. 
The initial data $\Psi_{\eta_0}$ is required to respect the semiclassical approximation, that is, the discrepancy from a unitary evolution shall be negligible at the initial time $\eta_{0}$, given by $\Psi_{\eta_0}(\phi)\lvert_{\partial U_0} \approx 0$.
The maximum of $ N_{\phi_0,\Psi_\eta} (\lvert \Phi(f) \rvert) $ also requires $U_0$ to be chosen as the largest possible neighborhood, for which $\phi_\Lambda(p) = m_p$ is the Planck mass.

While $\text{max}\left( N_{\phi_0,\Psi_t}( \mathcal R(\mathcal \lvert \Phi(f) \rvert )) \right) \equiv N_\text{max}^{ \lvert \Phi(f) \rvert } = 1$ marks undeniably the breakdown of the perturbative method, it is worth reminding ourselves that the semiclassical framework can become unreliable well before $N_\text{max}^{ \lvert \Phi(f) \rvert } =1 $.
This is because admissible fluctuations satisfying  \eqref{Filtration_for_EFT_Cosmology} do not imply that they respect the background configuration following the requirement of \eqref{cond_bs}.
As a consequence, the criterion by \eqref{cond_bs} must be determined from the underlying spacetime geometry, and always leading to an upper bound lower than $N_\text{max}^{ \lvert \Phi(f) \rvert }<1$.

\subsection{Random Field Simulation}

To connect our mathematical assessment of the viability of an effective framework in curved space-times with real data, we shift gears and present a Gaussian random field simulation\footnote{The source codes for the numerical experiments are available at \href{https://github.com/khchoi-lmu-physik/grf_qftcs}{\textcolor{blue}{\texttt{https://github.com/khchoi-lmu-physik/grf\_qftcs}}}.} to estimate  \eqref{Filtration_for_EFT_Cosmology}. 
We will demonstrate explicitly that the breakdown of the effective semiclassical framework in dynamic space-times goes hand in hand with unitarity loss.

While \eqref{Filtration_for_EFT_Cosmology} is  challenging to compute, 
it is conceptually straightforward because the spatial average of the expectation value of $\lvert \Phi(x) \rvert$ is associated with $\rho_\eta(\varphi)$ across all possible quantum field configurations within the neighborhood $U$.  

To navigate around computational complexity, we employ random field simulations \cite{hristopulos2020random, Bertschinger_2001} to estimate  \eqref{Filtration_for_EFT_Cosmology}.  
The random fields $\varphi_\text{sim}$  are designed to emulate the random variables $\varphi$ in $\rho_\eta(\varphi)$. 
This means that  $\varphi_\text{sim}$ mirrors the statistics of $\rho_\eta(\varphi)$ as in \eqref{gauss:mean}-\eqref{gauss:std}.
The only difference is that $\varphi_\text{sim}$ are discrete within the position space. Thus 
the resolution of the random fields is limited, i.e. they neither capture details finer than a pixel, nor macroscopic variations that are significantly larger than the simulation length scale. 
In technical terms, the spatial configurations are obtained through the inverse Fourier transform from its momentum space with a window function applied \cite{Mukhanov:1990me, Mukhanov:2005sc, Mukhanov:2007zz}.
This introduces a band-pass filter that blocks modes that are too long or too short in wavelength.  

\begin{figure*}[t!] 
  % Row 1 
    \includegraphics[width=0.4\linewidth]{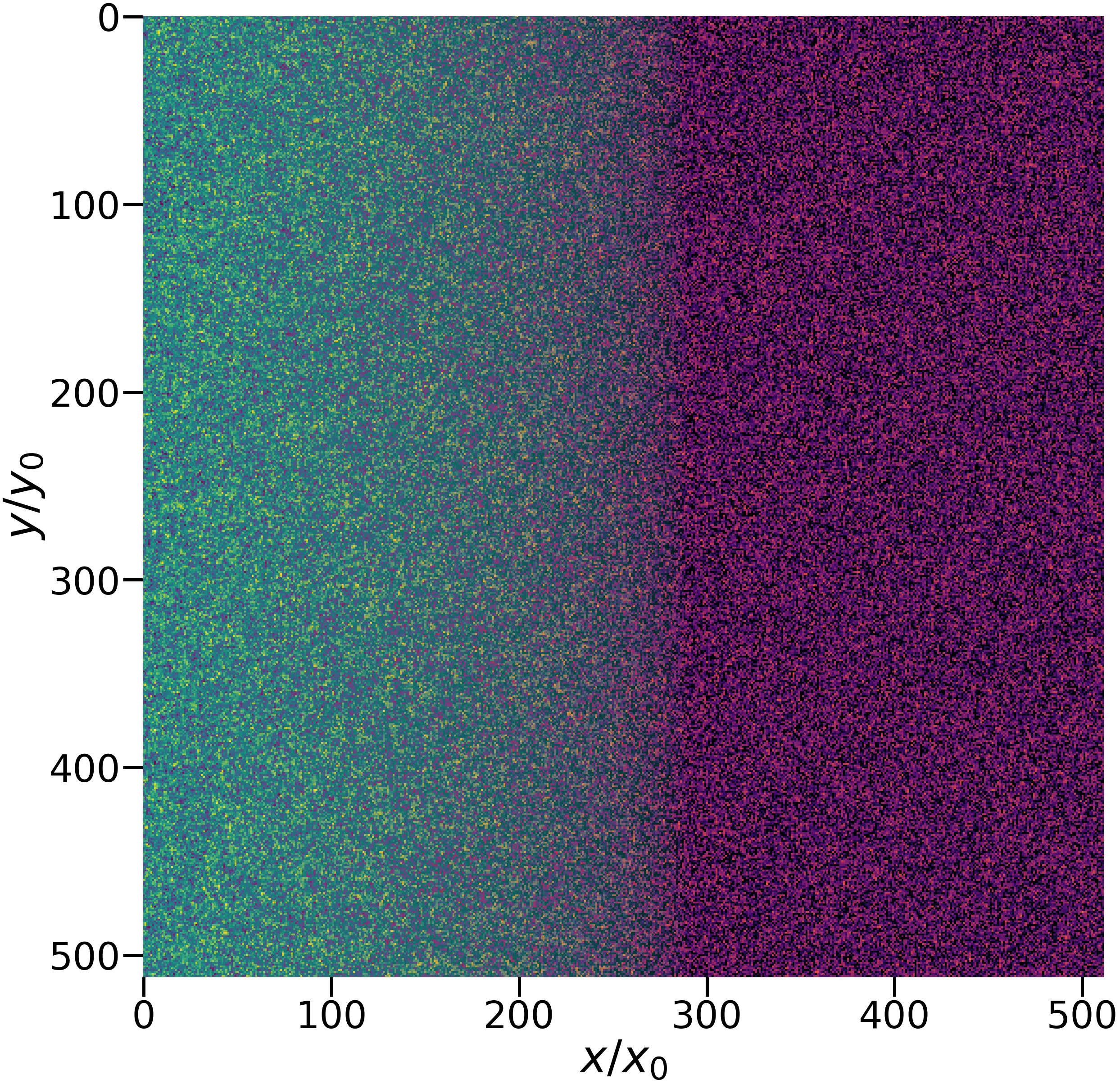}
    \includegraphics[width=0.4\linewidth]{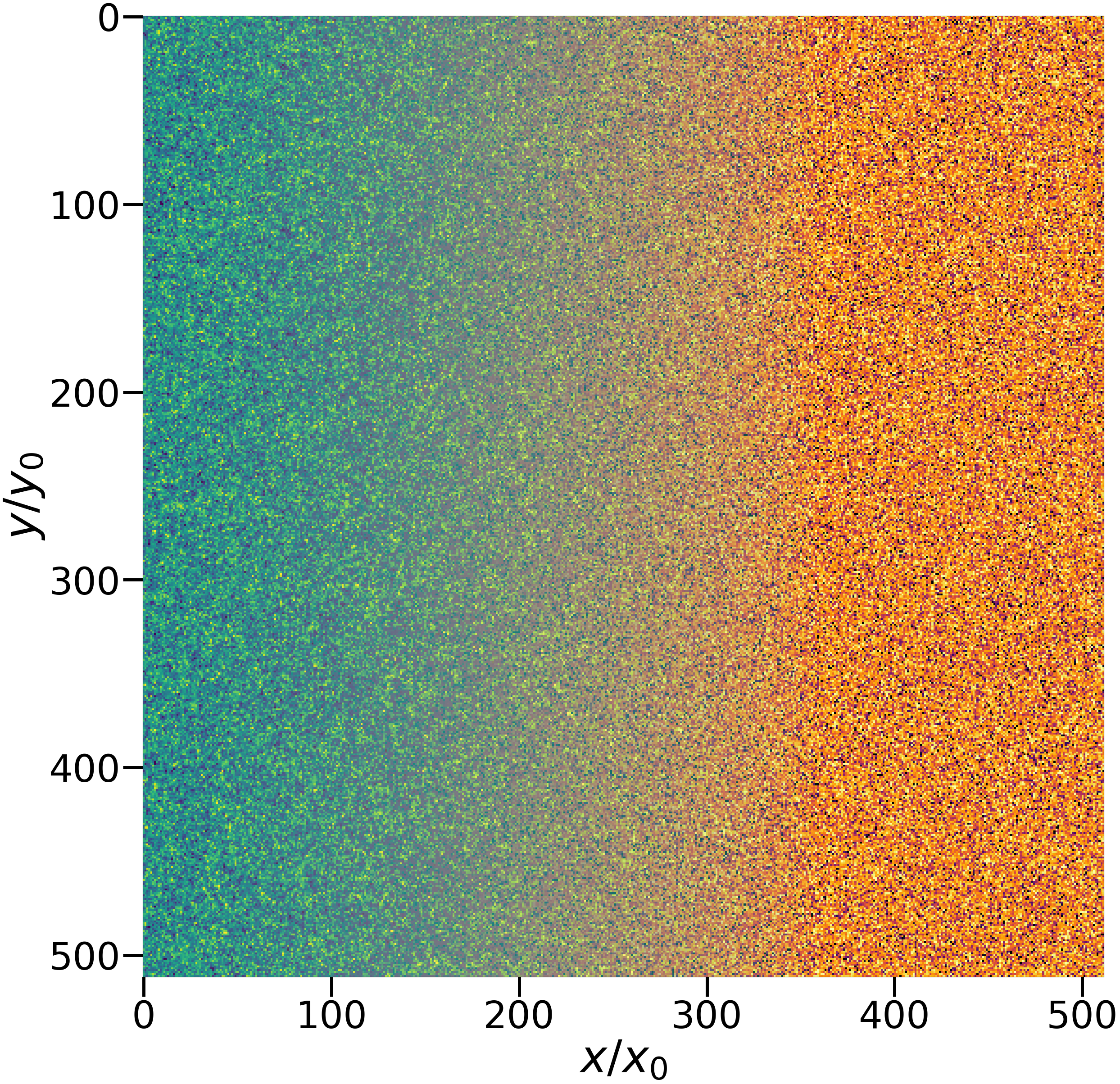}
    \quad 
    \includegraphics[width=0.08\linewidth]{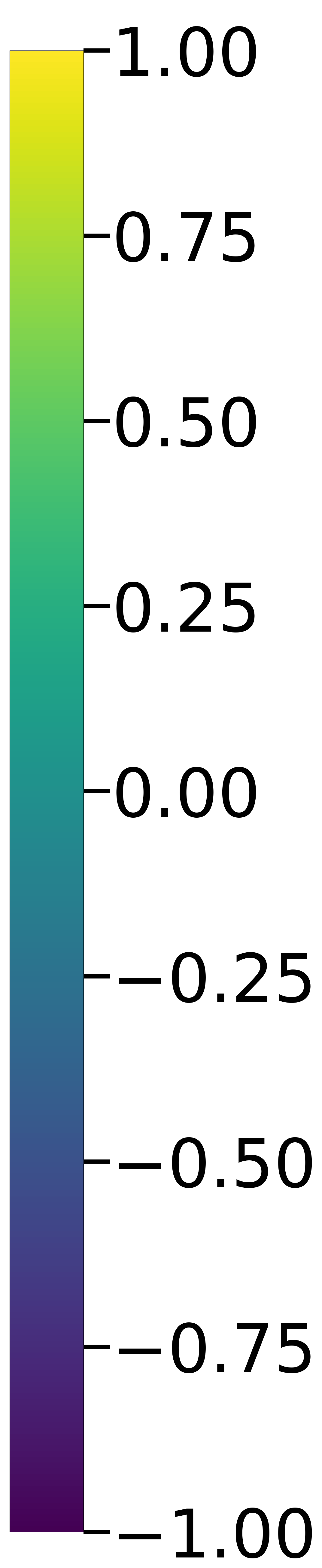}     
    \includegraphics[width=0.08\linewidth]{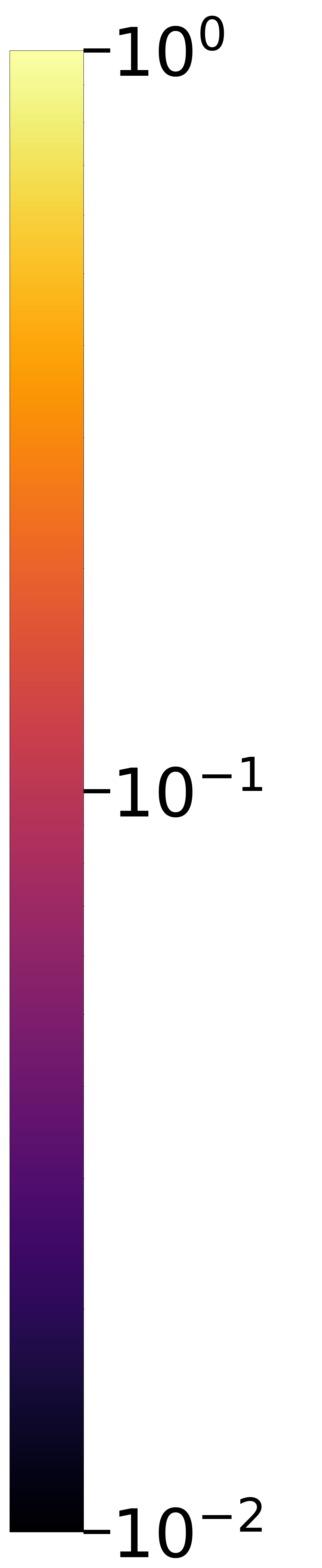}    
  \caption{Gaussian random field simulations of quantum fluctuations in a radiation-dominated universe  
  within a volume of scale $k_0^{-3}$.
  The random fields are evaluated on the $x$-$y$-plane at $z=1/2k_0$ at conformal times $\eta=10\eta_0$  (left) and $\eta=\eta_0$ (right).
  Each panel shows the normalized amplitude (left) and the magnitude (right). 
  As the system evolves into the high-curvature region from $10\eta_0$ to $\eta_0$, spacetimes dynamics amplify quantum fluctuations by at least two orders of magnitude without structure formation.
  Quantum fluctuations that are outside the domain of validity, $|\varphi_\text{sim}|>\varphi_\Lambda=1$, are excited everywhere and marked by the brightest spots at $\eta= \eta_0$. 
  This hints, at a preliminary level, at a substantial unitarity loss in our system at the time scale $\eta=\eta_0$, indicating the breakdown of the semiclassical approximation in the effective field theory.} 
  \label{fig:random_field_RDU}
\end{figure*}

\begin{figure*}[t!] 
  % Row 1 
    \includegraphics[width=0.43\linewidth]{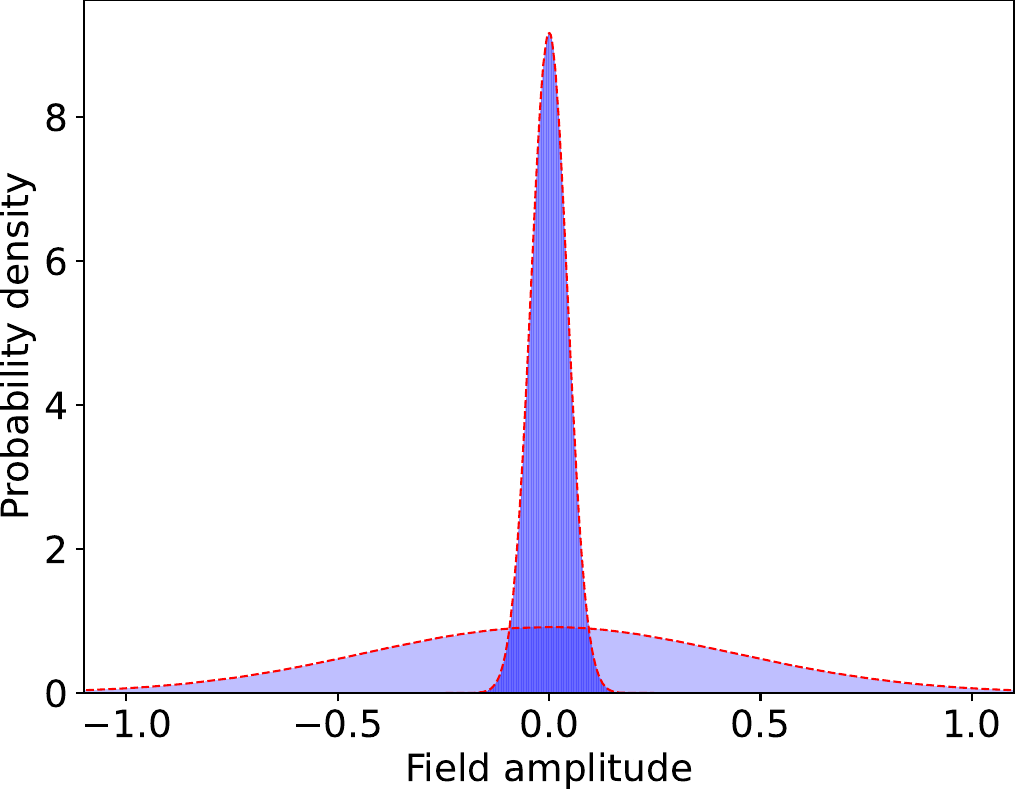}\quad \quad
    \includegraphics[width=0.45\linewidth]{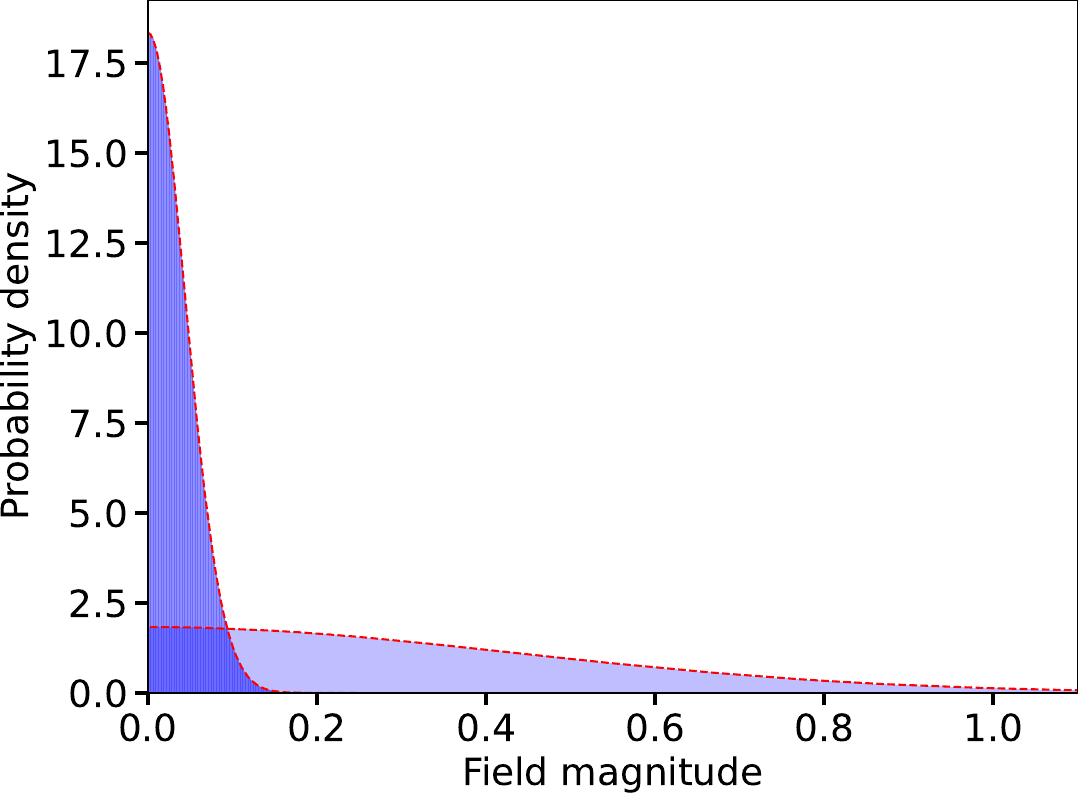} 
    \includegraphics[width=0.45\linewidth]{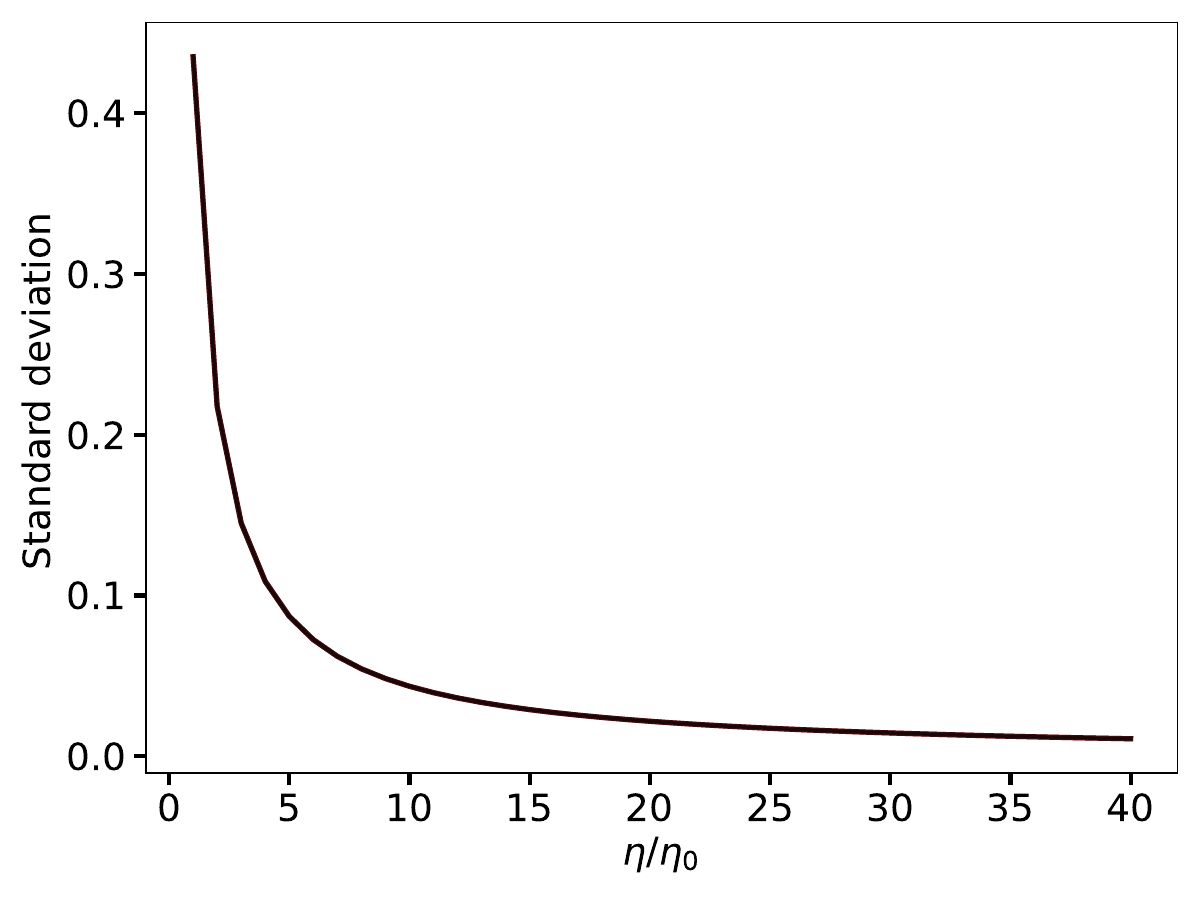}    
    \quad \quad
    \includegraphics[width=0.45\linewidth]{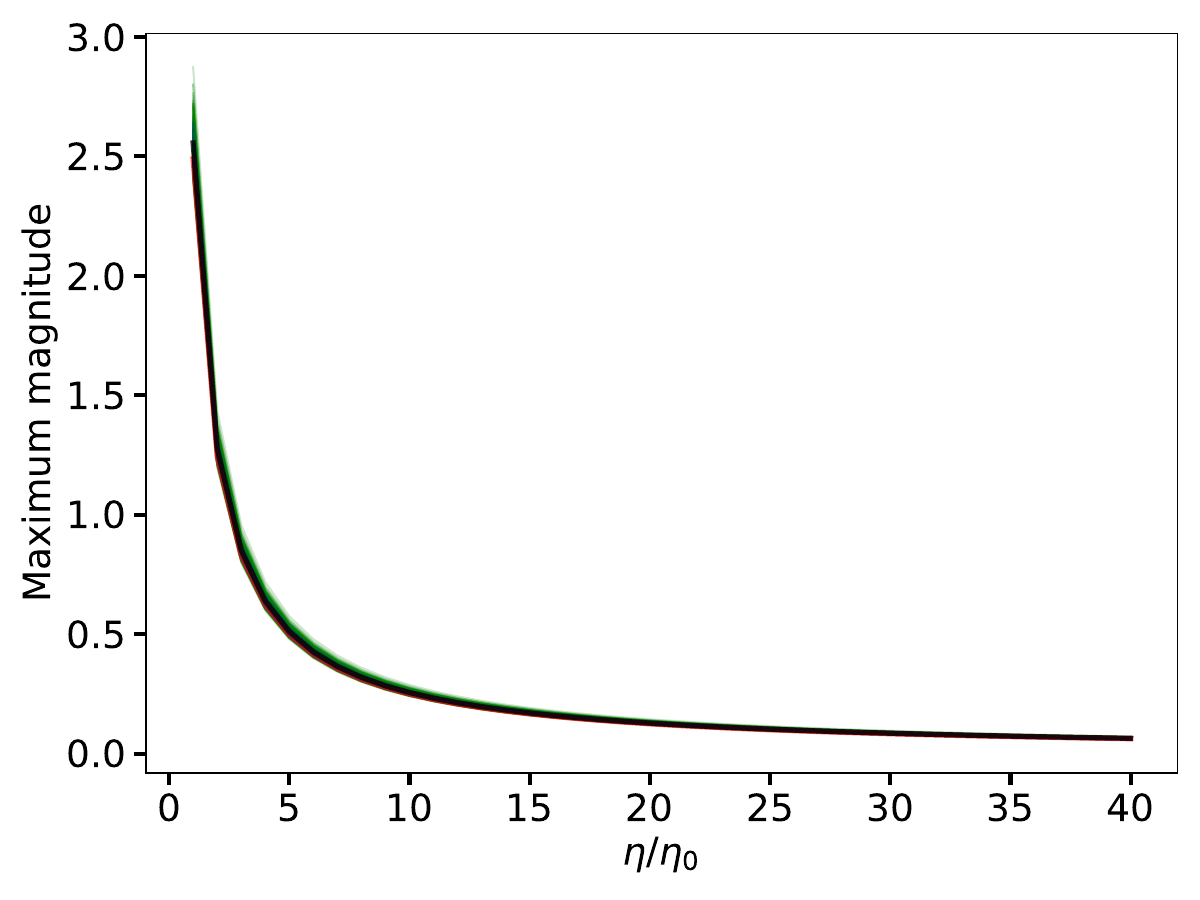}    
  \caption{Statistics of the Gaussian random field in the radiation-dominated universe. 
  The top row shows the probability density distribution of the random field in Fig. \ref{fig:random_field_RDU}  at conformal times $\eta= 10\eta_0$ and $\eta=\eta_0$. 
  The left panel shows the amplitude normalized by three times its standard deviation, while the right panel shows the magnitude.
  As the system evolves from $\eta= 10\eta_0$ to $\eta=\eta_0$, the initially narrow distribution flattens, indicating the amplification of quantum fluctuations when fields enter spacetime regions with intense dynamics. 
  This amplification causes a probability flux to leak beyond the domain of validity marked by the boundary $\lvert \varphi_\Lambda \rvert = 1$, signaling a strong unitarity loss as time progresses beyond $\eta=\eta_0$.
  The bottom row reinforces this interpretation by tracking the standard deviation of the random field and its maximum magnitude across $200$ simulations during the time interval $\eta/\eta_0 \in [1,40]$ with a step size of $\Delta \eta/\eta_0 = 1$.
  The red, green, and black curves represent the evolution in Fig. \ref{fig:random_field_RDU}, individual simulations, and their averages, respectively, and all are overlapping. 
  This demonstrates that the semiclassical framework breaks down at high confidence when $\eta\leq \eta_0$, since the standard deviation diverges consistently across all random field simulations.
  } \label{fig:RDU_statistics}  
\end{figure*}

To estimate \eqref{Filtration_for_EFT_Cosmology} via random field simulations, we calculate the spatial mean of $\lvert \varphi_\text{sim} \rvert$ for each simulation  
and then average this quantity across all simulations. 
Since $\varphi_\text{sim}$ follows the same Gaussian statistics as given by $\rho_\eta(\varphi)$,  $\lvert \varphi_\text{sim} \rvert$ follows a half-normal distribution (See Fig. \ref{fig:RDU_statistics} and \ref{fig:desitter_staistics} for example). 
This statistical relation indicates that the spatial average of $\lvert \varphi_\text{sim} \rvert$ is given by the standard deviation of  $\varphi_\text{sim}$, up to a factor of  $\sqrt{2/\pi}$. 
Using this relation, we can estimate \eqref{Filtration_for_EFT_Cosmology} based on the standard deviation of $\varphi_\text{sim}$ as shown in Fig. \ref{fig:RDU_statistics} and \ref{fig:desitter_staistics}, which is taken by averaging the statistics over a large sample of random fields.  

We will investigate two distinct cosmological scenarios: a contracting radiation-dominated universe and an expanding de Sitter spacetime. 
The former presents a straightforward example to demonstrate the breakdown of the semiclassical framework due to the amplification of quantum fluctuations\footnote{
Although our analysis focused on a radiation-dominated universe, it extends generally to any spacetime dynamics that significantly amplify quantum fluctuations, leading inevitably to the same conclusion.}. 

Consider a contracting radiation-dominated universe where $a(\eta) = a_0 \eta $ with $\eta \in (-\infty, 0)$, and set $a_0=1$. The mode $u_k(\eta)$ is given by
\begin{align}
    u_k(\eta) = \frac{1}{  \sqrt{2k}a_0\eta} e^{\mathrm{i}k\eta}
\end{align}
resulting in $\sigma^2(k,\eta) \propto 1/(k\eta^{2})$.
This scaling relation ($1/k$) indicates a consistent spatially correlated structure across all $\eta$, as shown in Fig. \ref{fig:random_field_RDU}.
In our simulation, we adopted a simple set of parameters\footnote{The specific values are chosen for convenience only. 
Our analysis can be generalized to any particular choice of $k_0$, $\eta_0$, or $m_p$.
The key point is the relationship between $k_0$, $\eta_0$, and $m_p$, which dictates the time scale when large-scale structures of relevant length scale $k_0^{-1}$ emerge, or when the effective framework breaks down.} by letting $k_0 = |\eta_0^{-1}| = 1$ and $m_p =1$.

Based on the above parameters, we will test the reliability of the semiclassical approximation using \eqref{cond_sd} and \eqref{cond_bs}. 
As a proof concept,  we compare $\lvert \Phi(f)\rvert$ with the Hubble parameter  $H_0= 1/\eta$, seleected as the classical observable for \eqref{cond_bs}.  
In a radiation-dominated universe, we expect \eqref{cond_bs} to hold consistently over time because $N_\text{max}^{|\Phi(f)|}$ scales with $\sigma \propto 1/\eta$, which matches the scaling of $H_0$.
This means that  \eqref{cond_bs} remains constant throughout the entire conformal time.

While \eqref{cond_bs} remains valid, \eqref{cond_sd} does not hold consistently. 
As shown in Fig. \ref{fig:random_field_RDU} and \ref{fig:RDU_statistics}, the contraction of the radiation-dominated universe significantly amplifies the magnitude of quantum fluctuations. 
This amplification can become so intense at late times that it drives a significant probability density current to leave the semiclassical domain with boundaries marked by $\lvert \phi_\Lambda \rvert =1$. 
This is clear in the flattening of the probability density distribution, as shown in  Fig. \ref{fig:RDU_statistics}. 

Additionally, Fig. \ref{fig:random_field_RDU} reveals that the quantum field at $\eta = \eta_0$ is populated with extreme signals everywhere, highlighted as the brightest spots representing $\lvert \varphi_\text{sim}\rvert > 1$.
When the system evolves beyond $\eta = \eta_0$, the semiclassical framework is expected to break down, as $\varphi_\text{sim}$ will will no longer satisfy the criterion \eqref{cond_sd}, where $ N_\text{max}^{ \lvert \Phi(f) \rvert }  < 1  $. 
The breakdown is even clearer in Fig. \ref{fig:RDU_statistics}, where the spatially averaged standard deviation of
 $\varphi_\text{sim}$ diverges as $\eta \rightarrow 0$. 
  
Consequently, the simulation at $\eta  = \eta_0$ should not be considered as an accurate description, but rather as giving the timing when the semiclassical effective framework breaks down.
 
Note that it is impossible to restore the condition $ N_\text{max}^{ \lvert \Phi(f) \rvert }  \ll 1$ at any later time $\eta_1$ by restricting ourself to a smaller neighborhood $U_1$. 
In fact, this only leads to more severe unitarity loss because a larger proportion of the probability will be excluded from the contraction of the neighborhood, hence further invalidating the existing framework.
Given that the standard deviation diverges as $1/\eta^2$ in Fig. \ref{fig:RDU_statistics} , we expect the following chain of inequalities upon contracting the neighborhood $U_2 \subseteq U_1\subseteq U_0$ for $\lvert \eta_1 \rvert \geq \lvert \eta_2 \rvert $: 
\begin{align}
    ...\leq 
    \|\Psi_{\eta_2}\|_{U_2} \leq 
   \|\Psi_{\eta_2}\|_{U_1} \leq 
   \|\Psi_{\eta_1}\|_{U_1} 
    \leq 
\|\Psi_{\eta_1
    }\|_{U_0} 
\end{align}
Ultimately, at some late time $\eta_\text{f} \rightarrow 0$, the only possible neighborhood is a set of measure zero with a zero norm. 

\begin{figure*}[t!]   
  % Row 1
  \flushleft 
       \includegraphics[width=0.21\linewidth]{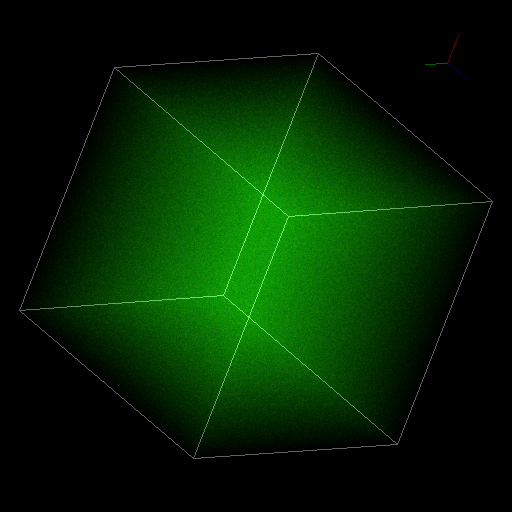}  
      \includegraphics[width=0.21\linewidth]{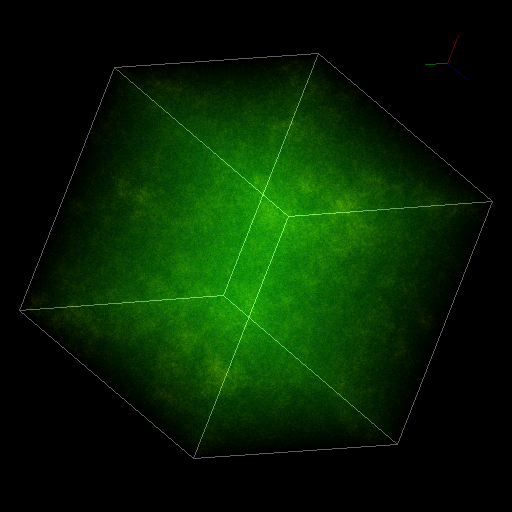}  
      \includegraphics[width=0.21\linewidth]{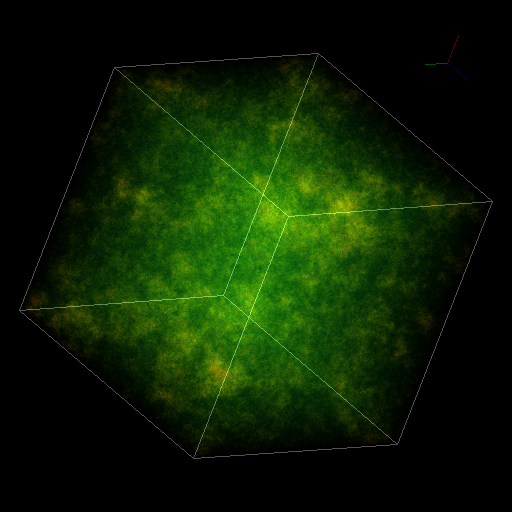}  
        \includegraphics[width=0.21\linewidth]{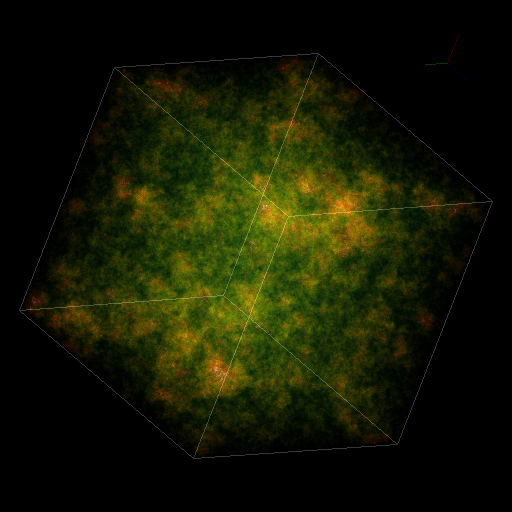}    
        \includegraphics[width=0.0325\linewidth]{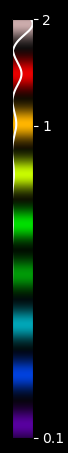}  
        \includegraphics[width=0.21\linewidth]{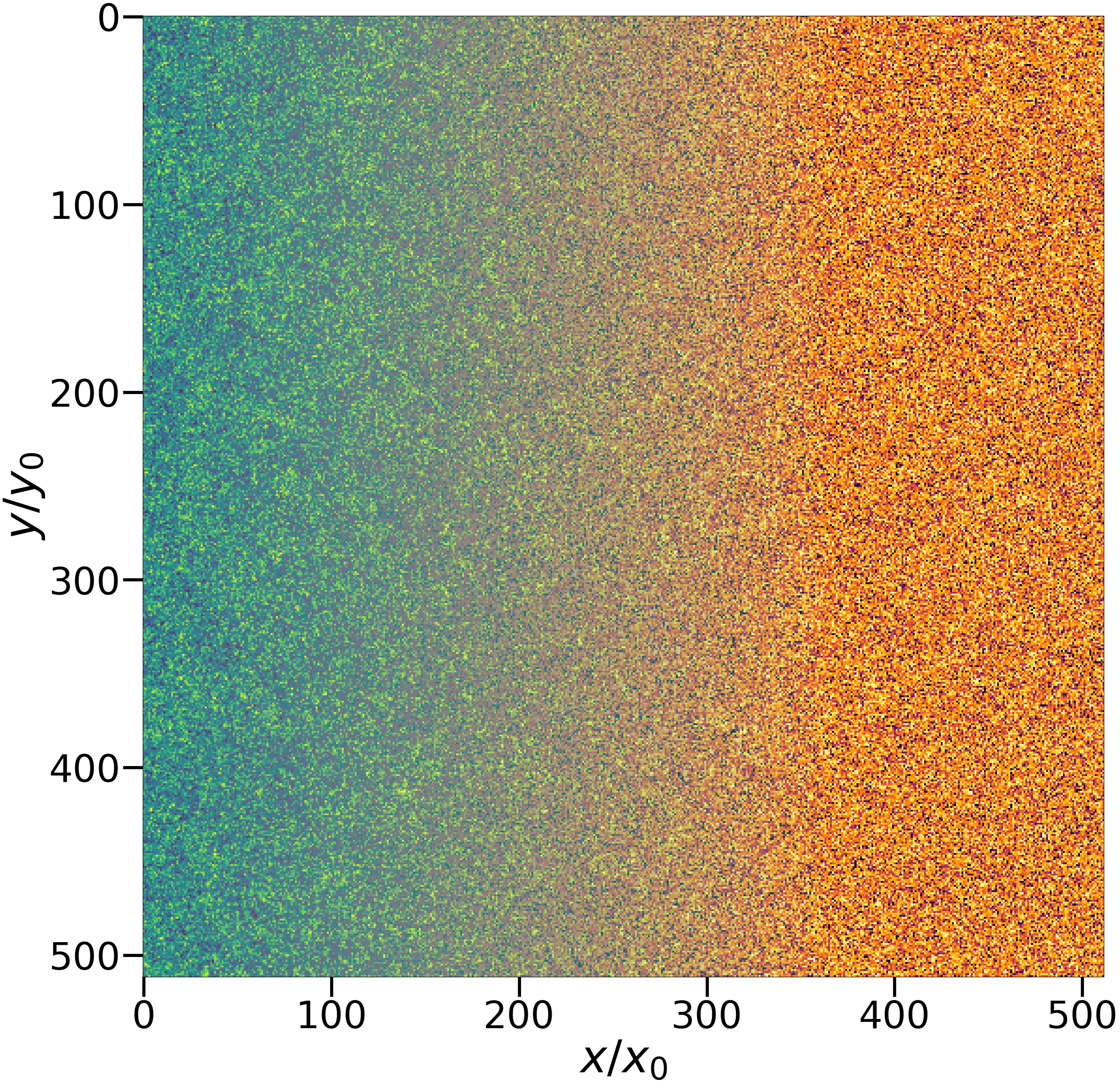}  
        \includegraphics[width=0.21\linewidth]{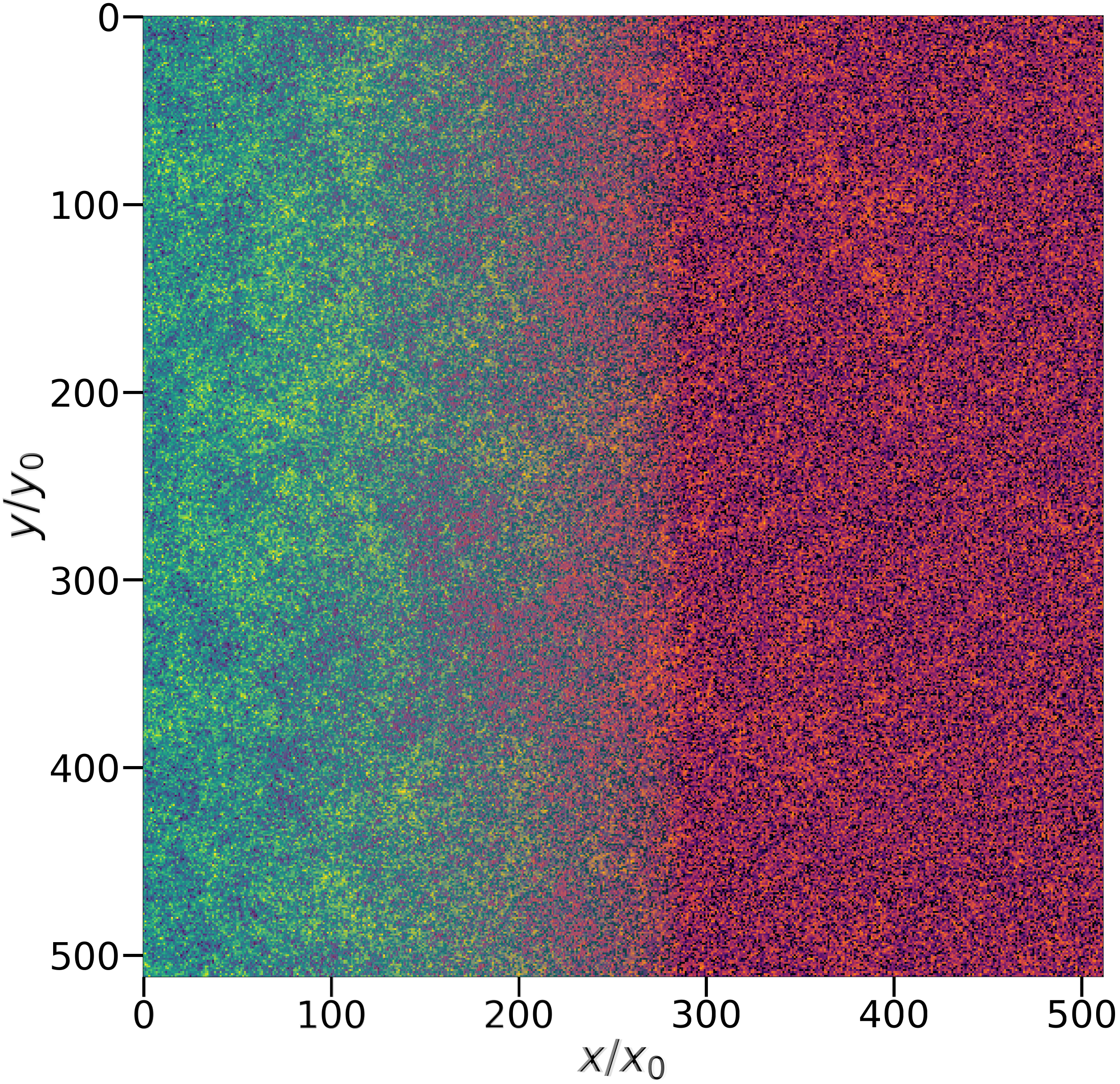}  
        \includegraphics[width=0.21\linewidth]{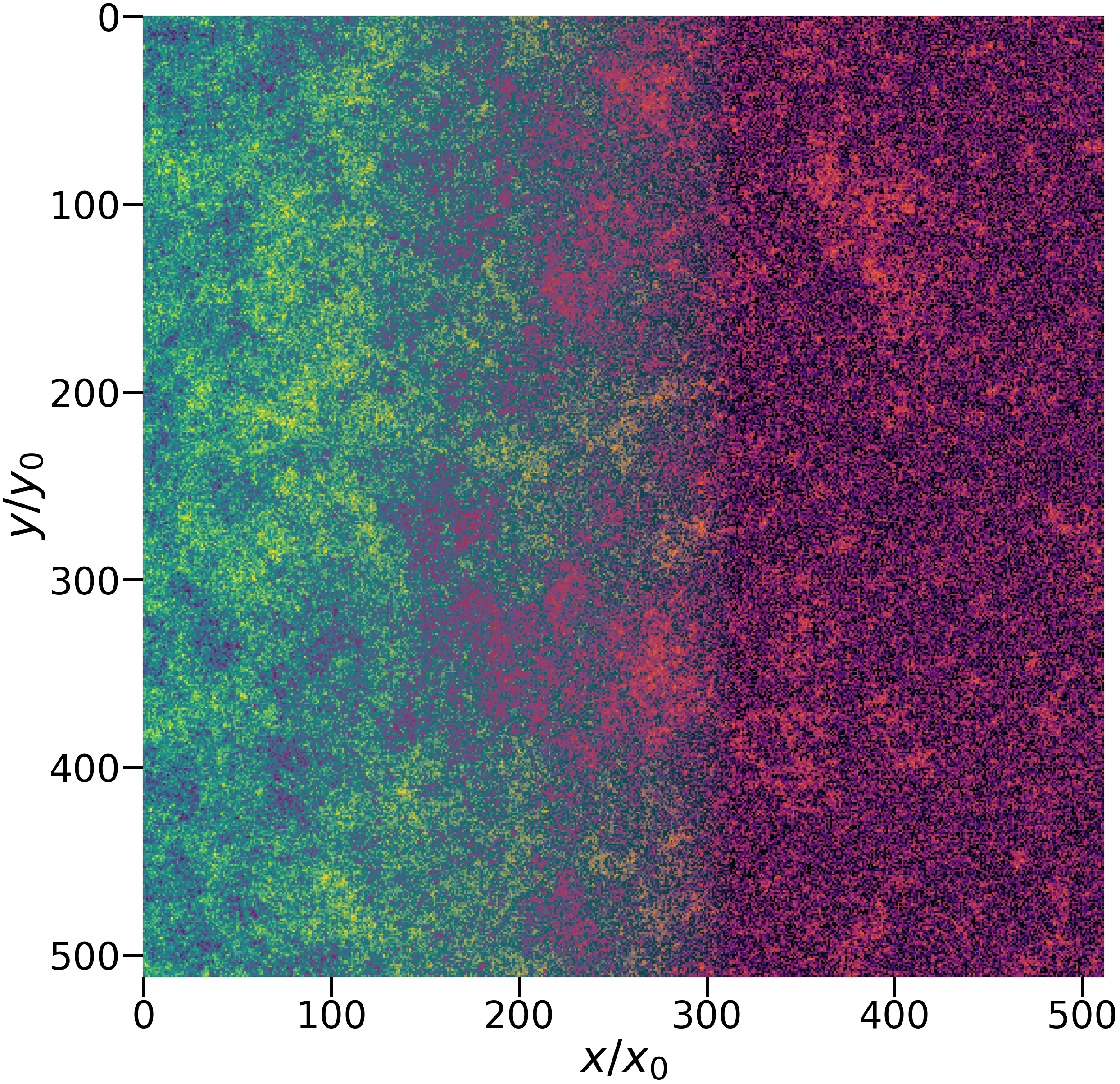}  
        \includegraphics[width=0.21\linewidth]{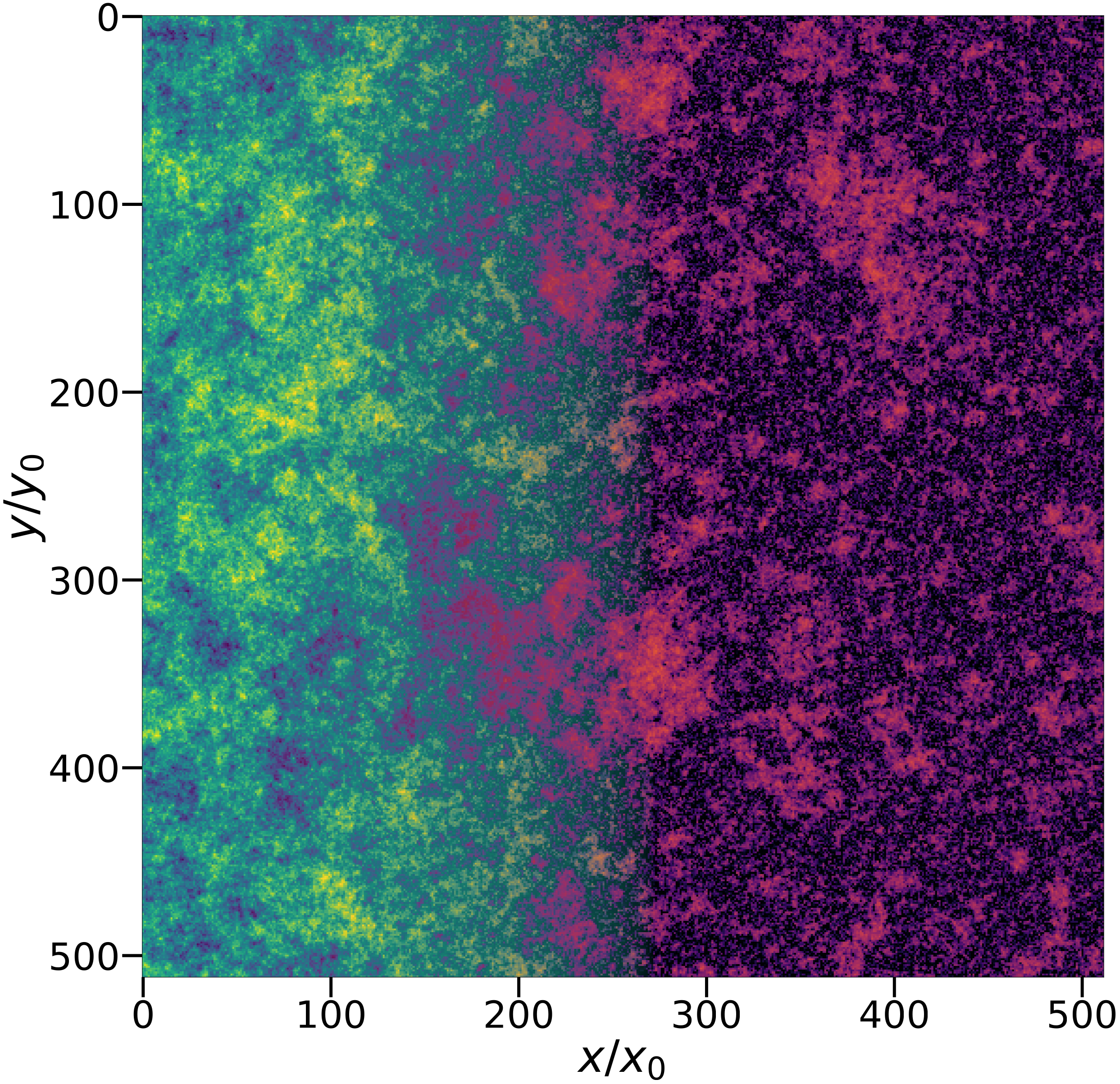}  
        \includegraphics[width=0.043\linewidth]{Figures/viridis_legend.png}   
        \includegraphics[width=0.043\linewidth]{Figures/inferno_legend.png}   
    \caption{Gaussian random field simulations of quantum fluctuations in the de Sitter spacetime within a volume of scale $ k_0^{-3}$ at conformal times $\eta/\eta_0= \{10, 2, 1, 0.2\}$ (from left to right, with $\eta<0$ and $\eta_0<0$ ).
    The top row displays the normalized magnitude of a three-dimensional volume-rendered random field, where stronger signals appear more opaque and weaker signals more transparent. 
    It demonstrates an initial homogeneous isotropic configuration at $\eta = 0.2 \eta_0$ developing large-scale structures as the system evolves into the future.
    The bottom row further illustrates this by showing the random field evaluated on the $x$-$y$-plane at $z=1/2k_0$ at the same conformal times.
    Each panel shows the normalized amplitude (left) and the magnitude (right). 
    In contrast to the radiation-dominated universe (Fig. \ref{fig:random_field_RDU}), as the system evolves forward in time, large-scale structures form while quantum fluctuations are diminished. 
    This indicates that any field configurations admissible in the far future originate from the distant past where extreme fluctuations are dominant, marked by the brightest spots.
    This suggests that, in order to comply with semiclassical approximation, any semiclassical effective framework must restrict its initial quantization to a smaller time domain. }  
  \label{fig:random_fields_desitter}
\end{figure*}  

In contrast to the collapsing radiation-dominated universe, the de Sitter spacetime requires a careful selection of initial data. 
The de Sitter spacetime is characterized by a scale factor $a(\eta) = (-a_\mathrm{dS}\eta)^{-1}$ for $\eta \in [-\infty, 0)$ and $a_\mathrm{dS}$ is the time-independent Hubble parameter, set to $a_\mathrm{dS} = 10^{-2}$ for convenient conceptual illustrations.
The modes $u_k(\eta)$ in de Sitter space-time are given by the Hankel functions 
\begin{equation}
    u_k(\eta)=a_\mathrm{dS} \frac{\sqrt{\pi}\eta^{3/2}}{2}\left( c_1H^{(1)}_\nu(k\eta)+  c_2 H^{(2)}_\nu(k\eta) \right)
\end{equation}
where $\nu^2 =\frac94 - 12(m^2/R+\xi)$. 
Adopting the Bunch-Davies vacuum where $c_1=0$ and $c_2= 1$ \cite{bunch1978quantum, Birrell:1982ix}, the variance becomes $\sigma^2(k,\eta) = \lvert a_\mathrm{dS}  (\sqrt{\pi}/2) \eta^{3/2} H^{(2)}_\nu(k\eta) \rvert^2$.   
For a minimally coupled massless scalar field ($\nu=3/2$), the Hankel function simplifies: 
\begin{equation}
    u_k(\eta)
    =\frac{\eta a_\mathrm{dS}e^{-\mathrm{i}k\eta}}{\sqrt{2k}}\left(1-\frac{\rm i}{k\eta}\right)
\end{equation}  
The explicit form indicates a transition in $\sigma^2(k,\eta)$ from $1/k$ to $1/k^3$ as $\eta$ evolves from $-\infty $ to $ 0$. This suggests the formation of large, spatially correlated structures, which is verified in Fig. \ref{fig:random_fields_desitter} illustrating that their formation effectively begins after $\eta   = 2 \eta_0$ (second panel from the left).  

Again we consider the criterion \eqref{cond_bs} where 
the Hubble parameter $\lvert H_0\rvert $ is chosen as the classical observable, such that the criterion \eqref{cond_bs} is given by $ \eta N_\text{max}^{ \lvert \Phi(f) \rvert } < \delta $, for $\delta=1/4$ as an example.

\begin{figure*}
  \centering
  % Row 1 
    \includegraphics[width=0.45\linewidth]{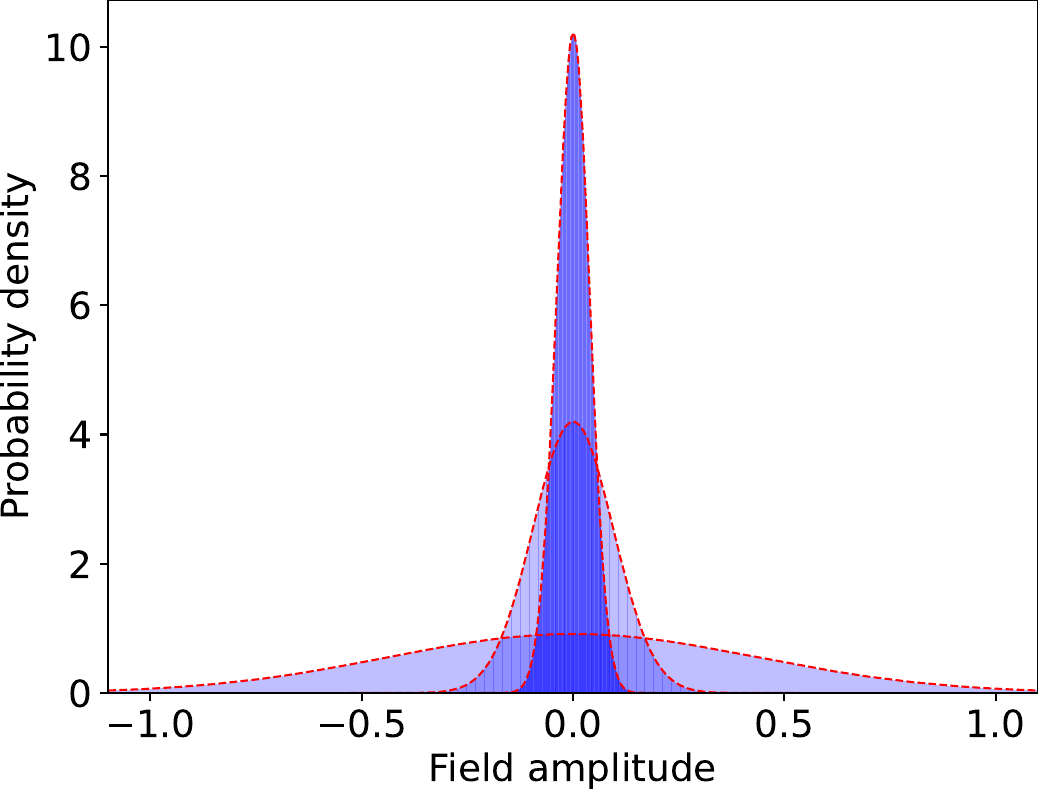}  
    \quad \includegraphics[width=0.47\linewidth]{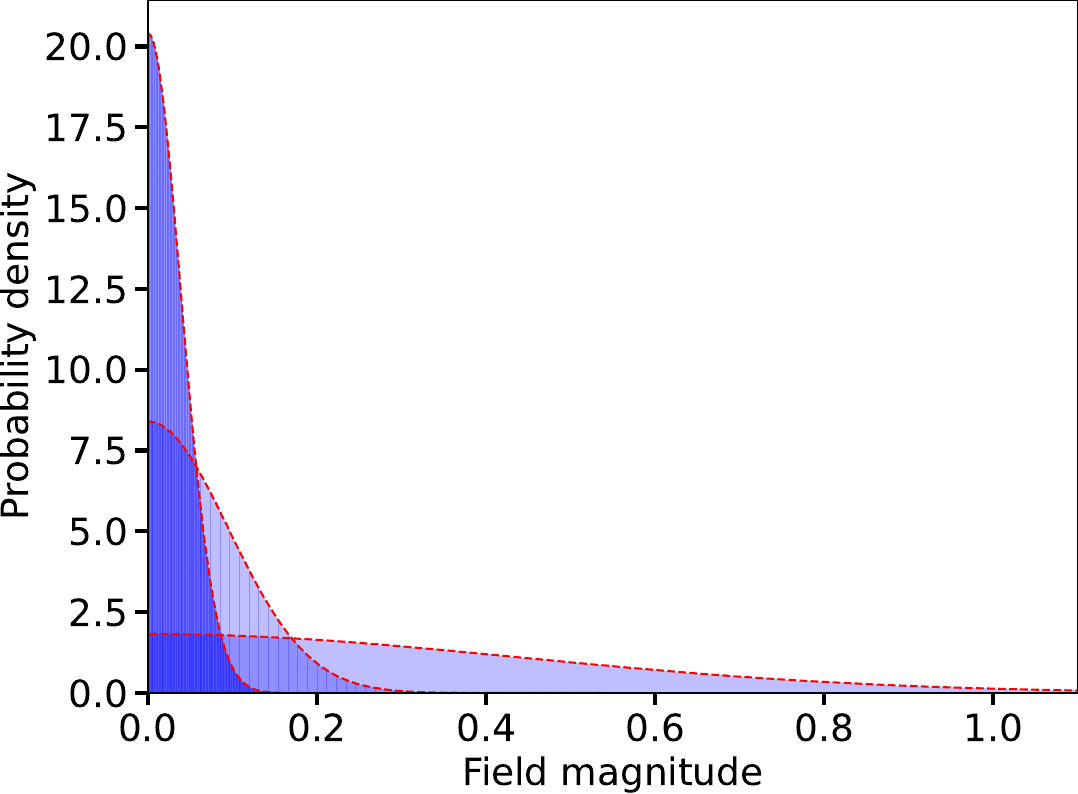}    
  \caption{ 
  Shown is the probability density distribution of the random field in Fig. \ref{fig:random_fields_desitter} at conformal times $\eta/\eta_0= \{10,2, 0.2\} $ (from flattest to steepest) in de Sitter spacetime. 
  The left panel shows the amplitude normalized by three times its standard deviation, while the right panel shows the magnitude.
  As the system evolves from $\eta= 10\eta_0$ to $\eta=0.2\eta_0$, the distribution bulges at its mean. 
  This indicates that the effective framework in an expanding de Sitter spacetime is highly reliable, with an approximately unitary evolution as the probability of exciting fluctuations beyond the semiclassical domain remains minimal for all late time.
  Nevertheless, although the effective framework remains robust in the future, the flattening of the distribution at the early time constrains the time domain for  initial quantization to be within the range of validity set by the consistency requirements \eqref{cond_bs} and \eqref{cond_sd}.} \label{fig:probability_density_desitter }
\end{figure*}

In strong contrast to a collapsing radiation-dominated universe, the expansion of the de Sitter universe suppresses the magnitude of quantum fluctuations.
This suppression is clearly visible in Fig. \ref{fig:random_fields_desitter}, and further demonstrated by the decrease in standard deviation over time as shown in Fig. \ref{fig:desitter_staistics}, alongside the narrowing of the probability distribution in Fig.  \ref{fig:probability_density_desitter } that indicates a low probability to excite fluctuations of the order of the short distance cut-off scale.

It follows from Fig. \ref{fig:desitter_staistics} that the condition $\eta  N_\text{max}^{ \lvert \Phi(f) \rvert }  < 1/4$ can be consistently fulfilled for all $\eta/\eta_0\leq 2$  as $\eta/\eta_0 \rightarrow 0 $.
However, the problem in de Sitter spacetimes will not arise at the late times, when fluctuations show greater variance in their magnitude as seen in Fig. \ref{fig:desitter_staistics}. In other words, fluctuations were more violent in the past.
Physically, this implies that while the field configurations are highly reliable at arbitrary late times, it is not sensible to rewind them arbitrarily far back into the past while relying on the semi-classical framework. 
This underlines the inconsistency of utilizing arbitrary initial data from the distant past, since such data fulfills neither \eqref{cond_sd}, nor \eqref{cond_bs}.
Therefore, only initial data after a specific time scale that respects the semiclassical assumption for a given length scale $1/k_0$ should be selected. 
Once equipped with valid initial conditions, the subsequent time evolution in de Sitter spacetime can be modeled under the effective semiclassical framework with excellent accuracy into the future.   

\begin{figure}[t!] 
  % Row 1 
  \centering\includegraphics[width=0.8\linewidth]{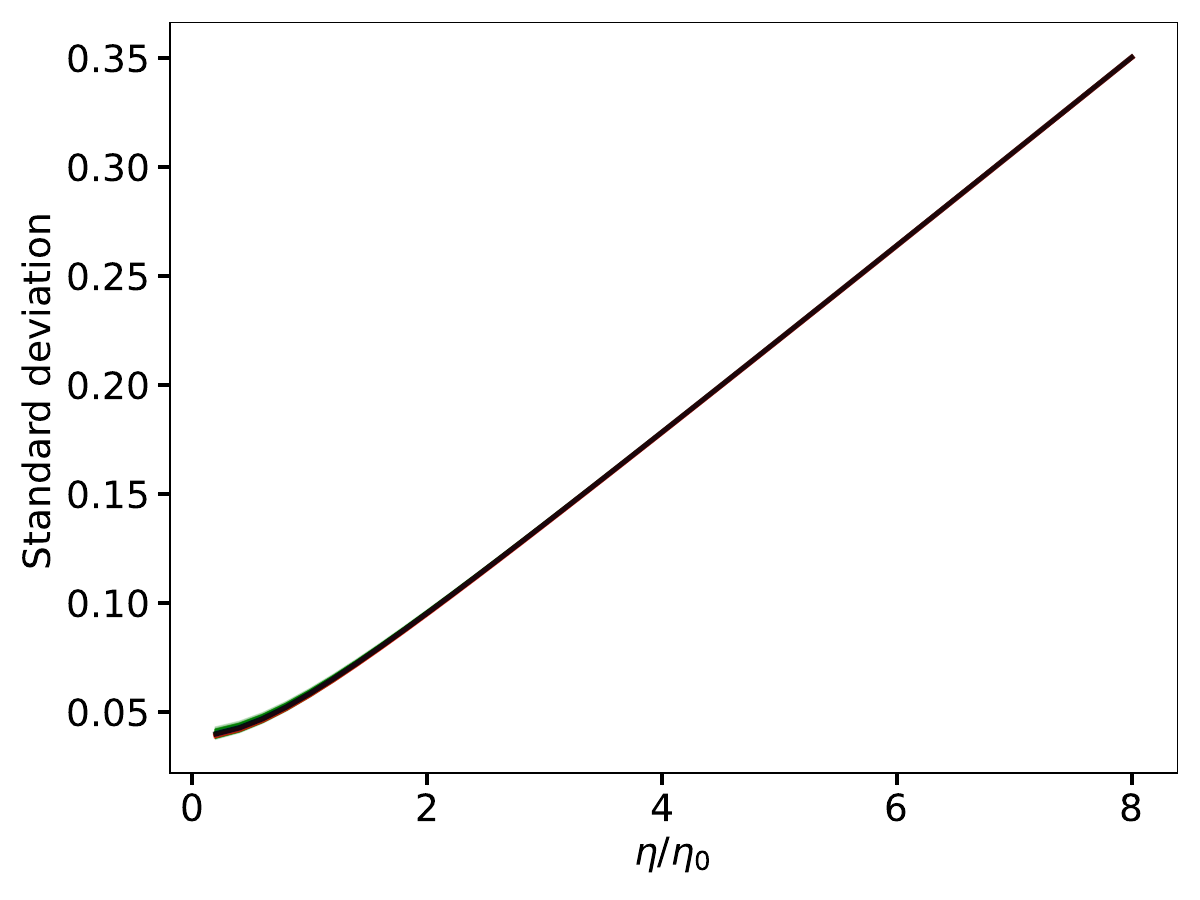}  
  \caption{Standard deviation of the probability density distribution in de Sitter spacetime across $200$ random field simulations during the conformal time interval $\eta/\eta_0 = [0.2, 8]$, with a time step of $\Delta\eta/\eta_0 =0.2$. 
  The red, green, and black curves represent the evolution of Fig. \ref{fig:random_fields_desitter}, individual simulations, and their averages, respectively, and are all overlapping. 
  As conformal time evolves backward into the past, the standard deviation scales linearly, with the consistency requirement \eqref{cond_bs} breaking down at $\eta\sim 3 \eta_0$ and \eqref{cond_sd} at $\eta\sim 25\eta_0$. 
  This indicates that fluctuations are extreme at early times, suggesting that not all initial times are suitable for setting initial conditions.}
   \label{fig:desitter_staistics}
\end{figure}

\begin{figure*}[t!]
\includegraphics[width=0.47\linewidth]{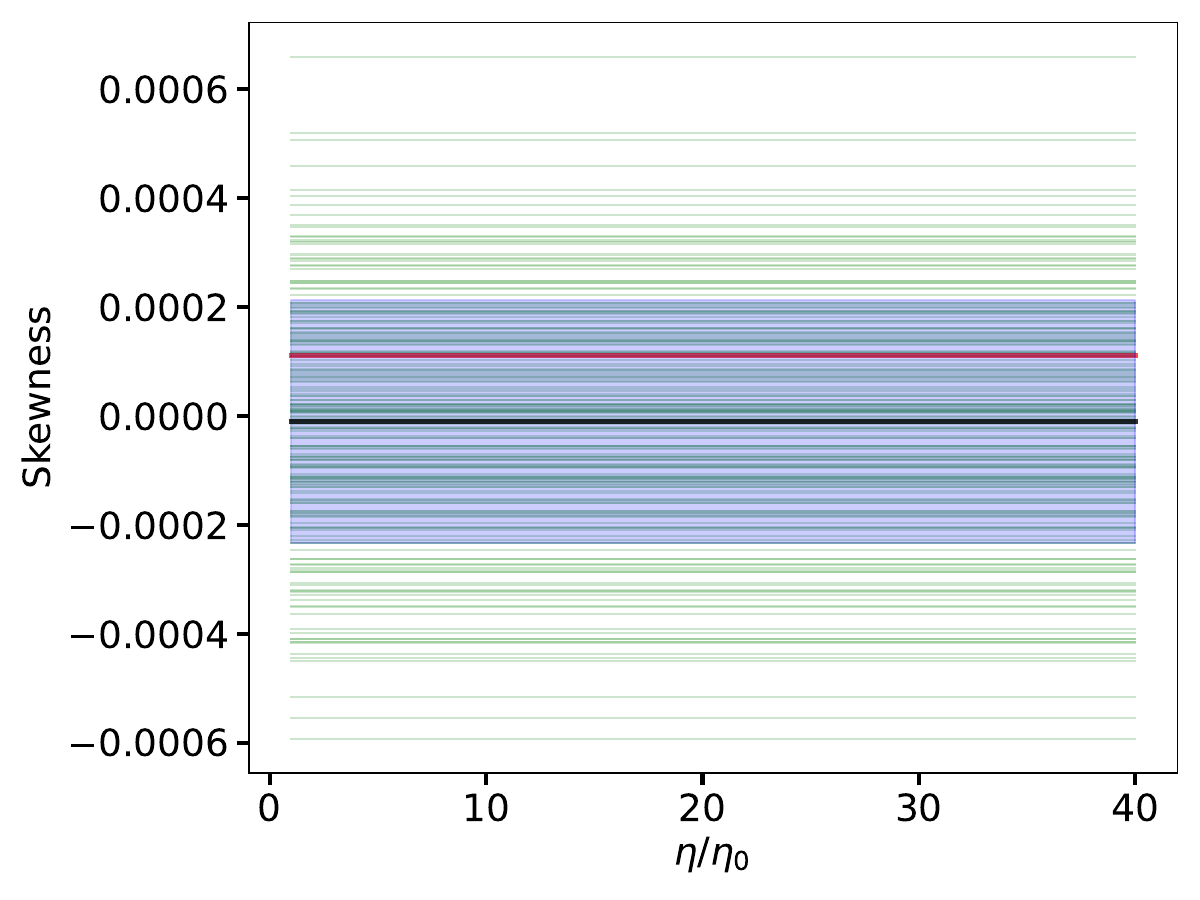}   
\quad 
\includegraphics[width=0.47\linewidth]{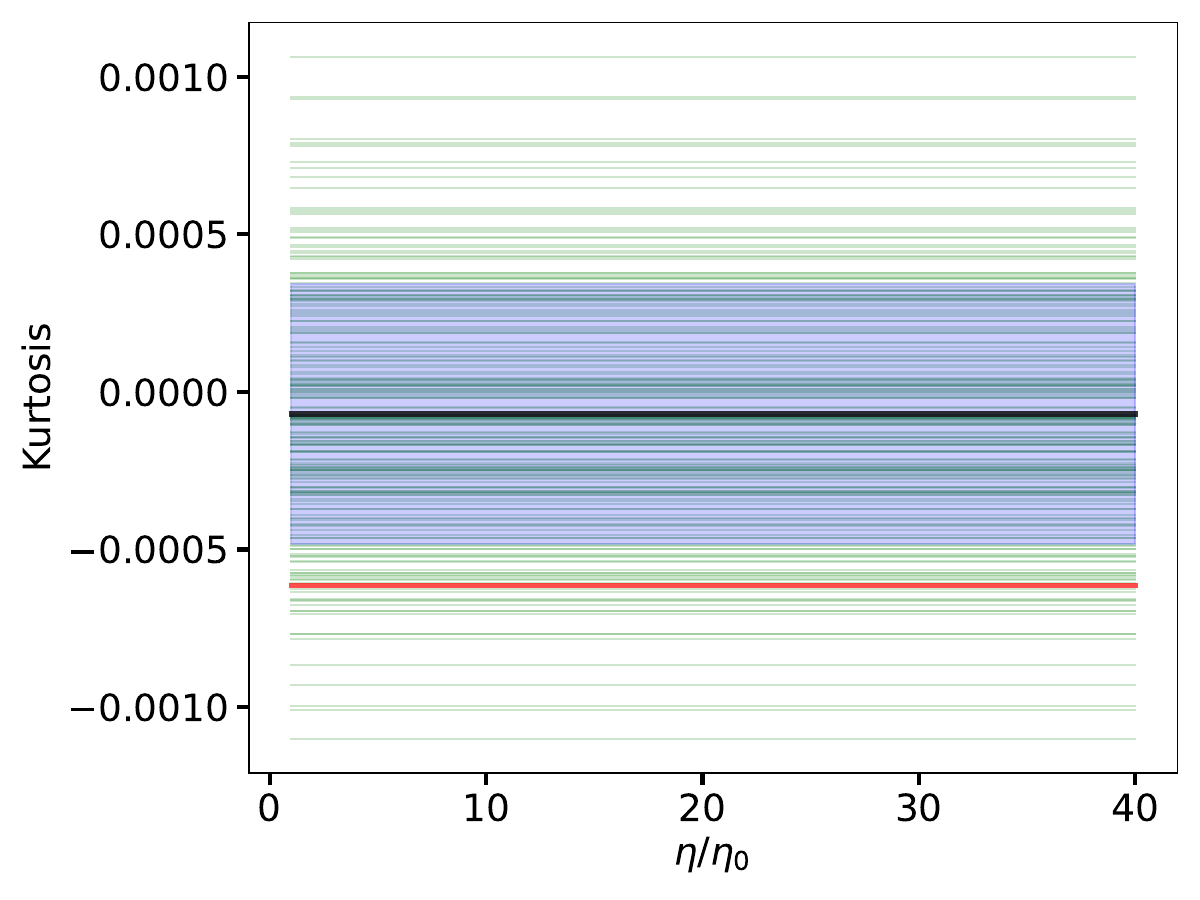}    \includegraphics[width=0.47\linewidth]{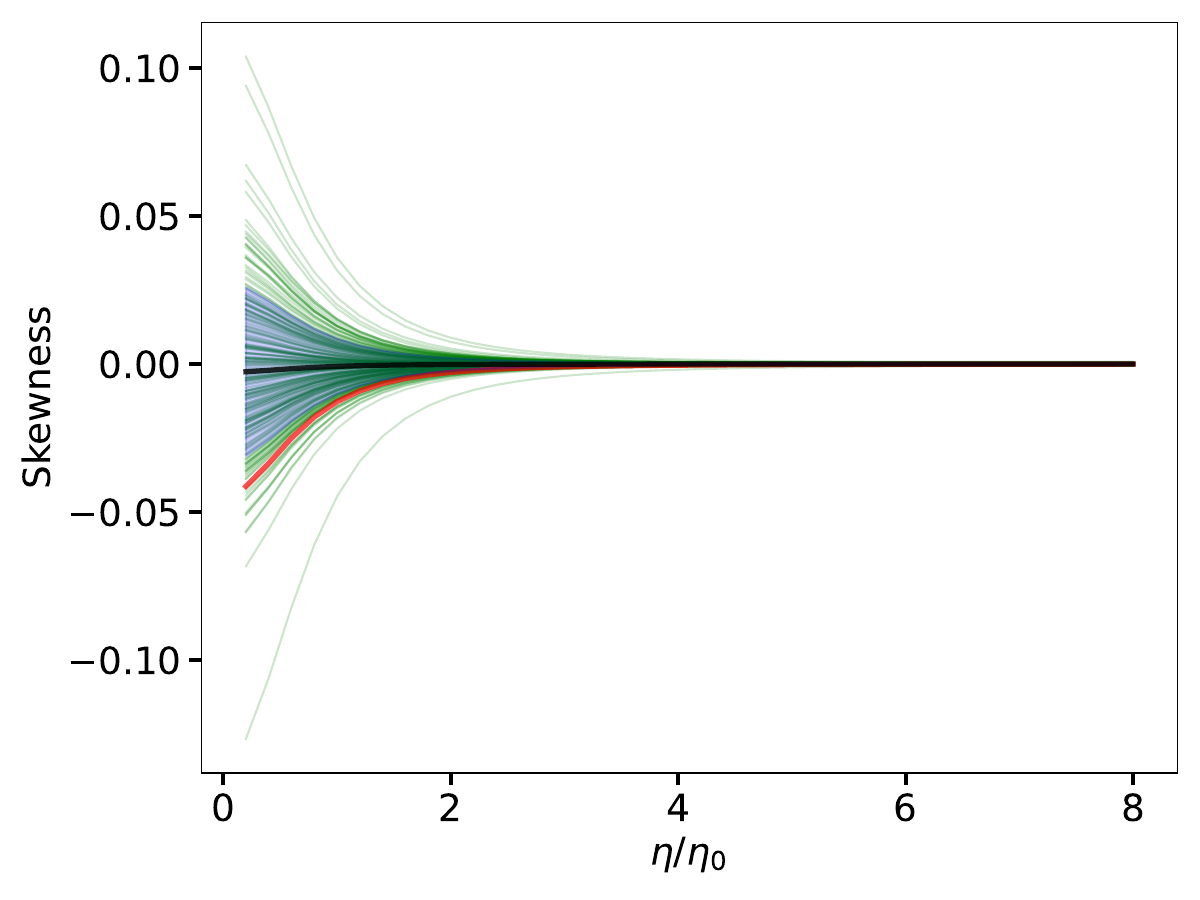}  
\quad 
\includegraphics[width=0.47\linewidth]{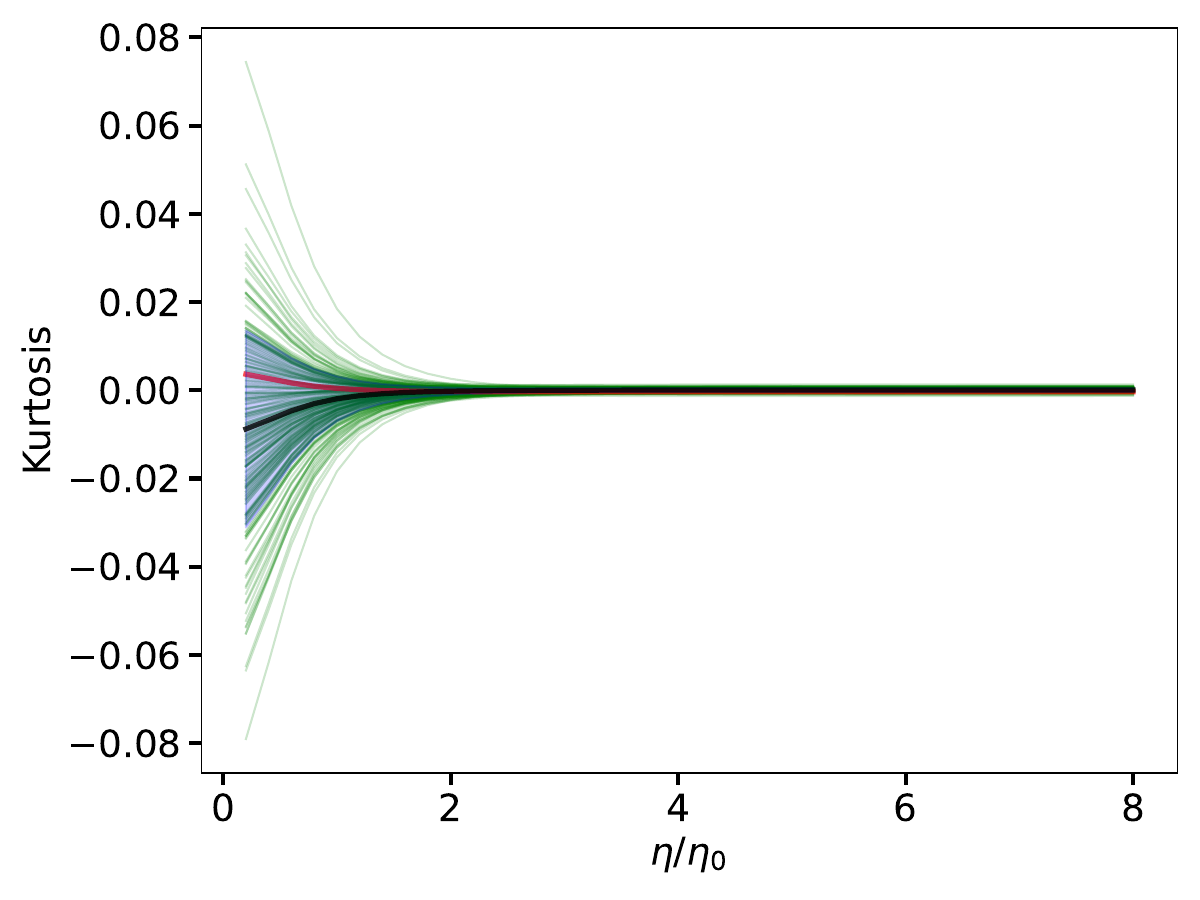}  
\caption{Normality test using kurtosis (right) and skewness (left) of the probability density distribution across 200 random field simulations in the radiation-dominated universe (top) and de Sitter spacetimes   (bottom).
  The red, green, and black curves represent the evolution of Fig. \ref{fig:random_field_RDU}, \ref{fig:random_fields_desitter}, individual simulations, and their averages, respectively. The light blue region highlights one standard deviation from the mean across all simulations.
  The qualitative difference observed is that in the radiation-dominated universe, higher moments remain constant and negligible; while in de Sitter spacetimes, these moments increase over time due to the formation of large-scale structures, as shown in Fig. \ref{fig:random_fields_desitter}.
  This increase in higher moments is caused by the finite data size used to describe strongly correlated data. 
  Nevertheless, this does not reject Gaussianity because we can see the absence of preferred higher moments indicated by the diminishing average of higher moments (black curve) as more simulations are sampled.}
   \label{fig:kurt_skew}
\end{figure*}

\section{Conclusion }\label{sec:conclude}

In this article we proposed a diagnostic framework to analyze the domain of validity of effective field theories in semi-classical spacetimes. 
The framework is based on local geometric requirements in configuration field space that address the effective short-distance incompleteness, as well as stability concerns related to the mean values of observables, which, in turn, restrict the semi-classical domain. 
This translates to boundaries in configuration field space associated with quantum fluctuations merely shifting the classical background configuration on scales considerably larger than the effective short-distance cut-off, or to boundaries associated with quantum fluctuations triggering sizeable backreactions on smaller scales.
In general, probability fluxes penetrating these boundaries lead to spacetime regions populated with quantum fluctuations that are outside the spectrum of fluctuations governed by the effective field theory. 
Therefore, dynamical aspects of the effective description are given by a contractive evolution (semi-) group, rather than a unitary representation of time translations, in order to account for the probability that is leaking into regions beyond the boundaries in field configuration space. 
The boundaries depend on the set of observables considered by the observers. 
Relative to this set the boundaries guarantee the consistency of the effective description, cf \eqref{cond_sd}, and the validity of the semiclassical approximation, cf \eqref{cond_bs}. 
In fact, \eqref{cond_bs} measures the magnitude of backreactions which is indicative for the background stability.

In order to quantize the effective field theory, we showed that the basic observables, in our case the configuration and its canonical momentum field, enjoy self-adjoint extensions on a specific Hilbert space.
At the kinematic level, the canonical momentum field operator admits infinitely many self-adjoint extensions. 
Via the usual functional calculus, the analysis extends to the Hamilton operator, again at the kinematic level.
However, at the dynamic level, in general, the solution space of the functional Schr\"odinger equation has no nontrivial intersection with the domains on which the Hamiltonian admits self-adjoint extensions. Important exceptions are static spacetimes. 
Therefore, provided that the theory is free of ghosts, the evolution must be contractive and the loss of unitarity, usually a sacrosanct requirement for a consistent (but also fundamental) probabilistic framework, has a well-understood reason: The existence of boundaries in field configuration space through which probability can be leaked.

Of course, the loss of unitarity is vital for the existence of a more fundamental description. There are spacetimes that when populated with quantum fluctuations leave the semiclassical approximation intact
because potentially harmful field excitations are too rare to impact the large-scale geometry. Intuitively these can be imagined as almost static or mildly dynamic spacetimes. Contrary to these there are spacetimes that support the excitation of free quantum fluctuations that trigger background instabilities and invalidate the semiclassical approximation at the global level.

In order to demonstrate the framework developed in this work, we complement our formal analysis by numerical experiments. As a proof of concept, with the aid of our numerical experiments, we identify regions in cosmological spacetimes, where freely evolving quantum fluctuations violate unitarity to an extent that the semiclassical approximation is locally invalidated.
As a concrete example, we consider a collapsing universe filled with radiation. 
This spacetime geometry borders on a future singularity. 
As can be expected, there is a substantial probability flux in its vicinity penetrating the configuration field boundaries, thereby exciting fluctuations that destabilize the background, leading to a breakdown of the semiclassical approximation (Fig.\ref{fig:RDU_statistics}). 
In contrast, and as another example, we consider quantum fluctuations populating de Sitter spacetime: In this case the semiclassical approximation holds towards the future (given appropriate initial conditions), but ceases to hold in the past. 
This indicates that semiclassicality further restricts acceptable initial conditions by limiting the time domain where initial quantization is performed. 
This is crucial because, even if regular initial conditions are chosen by hand, the semiclassical approximation can immediately break down once the system evolves outside the allowed time domain.

\acknowledgements

KHC thanks James Creswell for his expertise and discussions on topics related to Gaussian random fields.
The authors thank Maximilian Koegler for the initial collaboration on a related topic.
MS has been supported by the Italian Ministry of Education and Scientific Research (MIUR) under the grant PRIN MIUR 2017-MB8AEZ.
KHC has been financially supported by the Excellence Cluster ORIGINS.

\appendix 
 
\section{Notation and Conventions}\label{app:notation}

Throughout this article use the following convention and conventions: smeared operators $\mathcal O(f)$ are defined by $\mathcal O(f) \equiv \int_{\Sigma_t} \mathrm d^3x \, \sqrt{{\rm det}(q)}\, f(x) \, \mathcal O(x)  $, where $\Sigma_t$ denotes a spatial submanifold, $\sqrt{{\rm det}(q)}$ the metric determinant of $\Sigma_t$, and $f(x)\in\mathcal{C}_c^\infty(\Sigma_t)$ are smooth smearing functions of compactly supported on $\Sigma_t$. 
Integral signs are mostly expressed by $\int_x \equiv  \int_{\Sigma_t} \mathrm d^3x $ for spatial integral, and $\int_k \equiv \int_\mathbb{K} \mathrm d^3k/(2\pi)^3 $ for the momentum integral within some momentum space.
Concerning the metric, we will use mostly plus signature $(-,+,+,+)$ and we will work in natural units for the theoretical as well as the simulation part, if not stated differently. 

\section{Normality Test }\label{app:nt}

In this section, we examine the normality of the Gaussian random fields simulated in de Sitter spacetime and the radiation-dominated universe. Our statistical analysis focuses on the higher moments of the random fields, specifically the kurtosis and skewness.

Fig. \ref{fig:kurt_skew} reveals a strong contrast between the previously considered scenarios:  
in the radiation-dominated universe, the probability density distribution of the Gaussian random field shows negligible higher moments that remain constant under time-evolution, indicating minimal deviation from normality. 
Instead, in the de Sitter spacetime, the higher moments grow in time as large-scale structures begin to form. 
We can see from Fig. \ref{fig:kurt_skew} that the development of large-scale structure introduces a non-trivial skewness and kurtosis in the probability distributions (Green curve).

The underlying cause of this phenomenon appears to lie the highly correlated nature of our simulation data, combined with the limitations of data size.
In de Sitter space, the spatial correlation is characterized by a $k^{-3}$ scaling in the power spectrum which implies that a neighborhood region is likely to contain signals of similar amplitude.  
This effect is particularly significant in de Sitter geometry, where the correlation length is comparable to the full simulation scale (See Fig. \ref{fig:random_fields_desitter}).

Although higher moments develop in each individual simulation, the statistical average of skewness across all simulations is approximately zero, with a minor negative averaged excess of kurtosis (black curve in \ref{fig:kurt_skew}).
This is important because it shows that there are no preferred higher moments across all simulations.
As more data is sampled, this diminishing effect is expected to suppress higher moments further.

\bibliography{references}

\end{document}